\documentclass[aps,superscriptaddress,
preprint,%
reprint,%
floatfix
]{revtex4-1}

\usepackage{amsmath}
\usepackage{comment}
\usepackage{dcolumn}
\usepackage{xr}
\usepackage{chngcntr}
\usepackage[colorlinks=true, allcolors=blue]{hyperref}

\usepackage{filecontents}

\usepackage{placeins}
\usepackage{color}
\usepackage{graphicx}
\usepackage{epstopdf}

\usepackage[caption=false]{subfig}

\newcommand{\pten}[2]{${#1\times10^{#2}}$}

\newcommand{\kv}{{\bf k}}
\newcommand{\qv}{{\bf q}}

\begin{document}

\author{Alexander N. Rudenko}
\email{a.rudenko@science.ru.nl}
\affiliation{\mbox{Key Laboratory of Artificial Micro- and Nano-Structures of Ministry of Education and School of Physics and Technology,} Wuhan University, Wuhan 430072, China}
\affiliation{\mbox{Institute for Molecules and Materials, Radboud University, Heijendaalseweg 135, NL-6525 AJ Nijmegen, The Netherlands}}
\affiliation{\mbox{Department of Theoretical Physics and Applied Mathematics,
Ural Federal University, 620002 Ekaterinburg, Russia}}

\author{Shengjun Yuan}
\email{s.yuan@whu.edu.cn}
\affiliation{\mbox{Key Laboratory of Artificial Micro- and Nano-Structures of Ministry of Education and School of Physics and Technology,} Wuhan University, Wuhan 430072, China}
\affiliation{\mbox{Institute for Molecules and Materials, Radboud University, Heijendaalseweg 135, NL-6525 AJ Nijmegen, The Netherlands}}

\date{\today}
	
\title{{Electron-phonon interaction and zero-field charge carrier transport \\ in the nodal-line semimetal ZrSiS}
}

\begin{abstract} 
We study electron-phonon interaction and related transport properties of nodal-line semimetal ZrSiS using first-principles calculations.
We find that ZrSiS is characterized by a weak electron-phonon coupling on the order of 0.1, which is almost energy independent.
The main contribution to the electron-phonon coupling originates from long-wavelength optical phonons, causing no significant renormalization
of the electron spectral function. At the charge neutrality point, we find that electrons and holes provide a comparable contribution
to the scattering rate. The phonon-limited resistivity calculated within the Boltzmann transport theory is found to be strongly direction-dependent with 
the ratio between out-of-plane and in-plane directions being $\rho_{zz}/\rho_{xx}\sim 7.5$, mainly determined by the anisotropy of carrier velocities.
We estimate zero-field resistivity to be \mbox{$\rho_{xx}\approx12$ $\mu\Omega$\,cm} at 300 K, which is in good agreement with experimental data.
Relatively small resistivity in ZrSiS can be attributed to a combination of weak electron-phonon coupling and high carrier velocities.
\end{abstract}

\maketitle

\section{Introduction}

Over the past decade, a novel class of three-dimensional (3D) materials known as Dirac and Weyl semimetals has been discovered \cite{Young12,Liu14,Borisenko14,Xu15}. These materials are often referred to as 3D graphene analogs because
their electronic properties are similar to the properties of two-dimensional graphene \cite{Castro-Neto09,Katsnelson-Book}. The electronic structure of 3D semimetals allows for the existence of gapless electronic excitations that are 
symmetry protected \cite{Yang14}. Unlike conventional semimetals, 3D Dirac materials demonstrate linear dispersion of the electronic spectrum forming a conical shape, similar to graphene. This gives rise to a large variety of unusual 
properties, being the subject of intensive research \cite{Burkov16,Armitage18}.

A special case of Dirac semimetals is the so-called nodal-line Dirac semimetals \cite{Fang15}, where the set of Dirac points forms a continuous (nodal) line in the electronic band structure. Nodal-line semimetals
host a large density of Dirac fermions that are favorable for the appearance of electron correlation effects \cite{Liu17}. Typical representatives of this class of materials are zirconium-based layered compounds with the chemical formula
ZrSi$X$, where $X$ is a group-VI element: S, Se, or Te. Among the ZrSi$X$ family of compounds, ZrSiS is an especially prospective material with a wide energy range of linearly dispersing bands reaching 2 eV. This makes this material an 
excellent candidate for studying Dirac fermions. ZrSiS has been extensively studied experimentally, demonstrating exotic physics and a rich spectrum of remarkable 
properties \cite{Schoop16,Neupane16,Wang16,Ali16,Singha17,Matusiak17,Sankar17,Lodge17,Butler17,Topp17,Hu17,Schilling17,Pezzini18}. Among them are the strongly nested Fermi surface \cite{Schoop16,Neupane16}, strong Zeeman splitting \cite{Hu17},
topologically nontrivial states \cite{Wang16,Ali16,Singha17}, high carrier mobility \cite{Matusiak17,Sankar17}, frequency-independent optical conductivity \cite{Schilling17}, and unconventional mass enhancement of quasiparticles near the
nodal line as indicated by quantum oscillations measured in high magnetic fields \cite{Pezzini18}. The latter finding suggests the importance of many-body effects, which may contribute to the realization of exotic quantum states in this
material. Particularly, it has been predicted that ZrSiS is a candidate for the formation of excitonic pairing \cite{Rudenko18}, whereas its counterpart ZrSiSe is predicted to host unconventional ($d$-wave) superconductivity \cite{Scherer18}.

Despite a large number of experimental studies on ZrSiS, which include angle-resolved photoemission spectroscopy, scanning tunneling microscopy, optical probes, and high-field magnetotransport measurements, physical mechanisms behind the
observable properties are still largely unclear. Moreover, some experimental studies report conflicting results, for instance, on zero-field resistivity \cite{Sankar17,Hu17,VanGennep19,Novak19}, which is an issue to be resolved.
Theoretical description of ZrSiS available in the literature is mostly limited to ground-state electronic structure calculations within density functional theory. Only a little
attention from the theory side has been given to the problem of unusual optical properties \cite{Habe18,Zhou} and many-body effects in ZrSiS \cite{Rudenko18,Scherer18}. The nature of intrinsic transport properties as well as mechanisms
responsible for high carrier mobility are not fully understood. The problem of lattice dynamics and electron-phonon coupling, which is supposed to be the main factor determining charge carrier scattering at room temperature, has only
scarcely been addressed \cite{Duman16,Zhou17,Singha18,Xue19}. At the same time, the scattering behavior (e.g., its doping dependence) in nodal-line semimetals is expected to be essentially different from that in conventional (semi-)conductors \cite{Syzranov17}. An 
in-depth microscopic understanding of the origin of the electron scattering and related phenomena in nodal-line semimetals constitutes an important step toward practical utilization of these materials.

In this paper, we perform a systematic first-principles study of electron-phonon interactions and analyze its influence on the transport properties in bulk ZrSiS. We find that ZrSiS is characterized by a weak electron-phonon interaction with the
coupling constant $\lambda\sim 0.1$, being almost independent of charge doping. The dominant contribution to the coupling originates from long-wavelength optical phonons. The electronic spectrum undergoes no significant renormalization
in the presence of electron-phonon interaction and is only distinguished by a finite linewidth. Around the charge neutrality point, electrons and holes in ZrSiS show similar contribution to the scattering rate, which exhibits
a pronounced V shape as a function of energy. The phonon-limited resistivities calculated within the semiclassical Boltzmann theory demonstrate strong anisotropy between the in-plane and the out-of-plane crystallographic directions with the 
ratio $\rho_{zz}/\rho_{xx}\sim 6.3$--$8.8$ depending on the doping level. Transport anisotropy is attributed to an essentially anisotropic carrier velocities rather than to the anisotropic electron-phonon scattering. The in-plane resistivity is estimated
to be $\rho_{xx}\approx 12$ $\mu\Omega$\,cm at 300 K, being in good agreement with experimentally measured values of 14--16 $\mu\Omega$\,cm \cite{Ali16,Singha17,Sankar17,Shirer19}. Relatively small resistivity can be attributed to weak electron-phonon coupling and 
high carrier velocities in ZrSiS.

The rest of the paper is organized as follows. In Sec.~\ref{sec2}, we describe basic theory and provide computational details. In Sec.~\ref{sec3}, our results and a related discussion are presented. In Sec.~\ref{sec4}, we summarize our findings and conclude the paper.

\section{Theory and computational details}{\label{sec2}}

In the semiclassical Boltzmann theory  \cite{Ziman,Grimvall}, the $\alpha\alpha$ component of the conductivity tensor has the form
\begin{equation}
    \sigma_{\alpha\alpha} = - \frac{e^2}{V}\sum_{n{\bf k}} \tau_{n{\bf k}} [v^{\alpha}_{n\bf k}]^2 \frac{\partial f_{n{\bf k}}}{\partial \varepsilon_{{n\bf k}}},
    \label{sigma}
\end{equation}
where $V$ is the unit cell volume, $v^{\alpha}_{n{\bf k}}$ is the $\alpha$ component of the group velocity for band $n$ and wave-vector {\bf k},  $\varepsilon_{n\bf k}$ is the corresponding electron energy, $f_{n{\bf k}}=\{1+\mathrm{exp}[(\varepsilon_{n{\bf k}}-\varepsilon_F)/k_BT])\}^{-1}$ is the electron occupation factor, and 
$\tau_{n\bf k}$ is the momentum-dependent scattering relaxation time. The latter can be related to the inverse linewidth of band $n$ at wave-vector ${\bf k}$,
\begin{equation}
	\frac{1}{\tau_{n\bf k}} = \frac{2}{\hbar}\mathrm{Im}(\Sigma_{n\bf k}),
\label{tau}
\end{equation}
which is expressed here via the imaginary part of the electron self-energy $\mathrm{Im}(\Sigma_{n\bf k})$ and known as the self-energy relaxation time approximation \cite{Ponce2018}. In the presence of electron-phonon coupling, the self-energy can be calculated within the Migdal approximation  \cite{Migdal1958}, leading to the following expression on the real frequency axis \cite{Marzari2012}:
\begin{multline}
        \Sigma_{n\bf k}(\omega,T)=\sum_{m\nu}\sum_{\bf q}|g_{mn,\nu}({\bf k},{\bf q})|^2 \\
        \times \left[ \frac{n_{{\bf q}\nu}(T)+f_{m{\bf k}+{\bf q}}(T)}{\omega-(\varepsilon_{m{\bf k}+{\bf q}}-\varepsilon_F)+\omega_{{\bf q}\nu}+i\delta} \right. \\
        + \left. \frac{n_{{\bf q}\nu}(T)+1-f_{m{\bf k}+{\bf q}}(T)}{\omega-(\varepsilon_{m{\bf k}+{\bf q}}-\varepsilon_F)-\omega_{{\bf q}\nu} + i\delta}   \right],
	\label{selfenergy}
\end{multline}
where $n_{\bf q\nu}$ is the occupation factor of a phonon with wave-vector ${\bf q}$, branch index $\nu$, and frequency $\omega_{{\bf q}\nu}$, and
\begin{equation}
g_{mn,\nu}({\bf k},{\bf q}) =  \bigg( \frac{\hbar}{2m_0\omega_{{\bf q}\nu} }
\bigg)^{1/2} 
\langle \psi_{m{\bf k+q}} | \partial_{{\bf q}\nu}V | \psi_{n{\bf k}}\rangle
\label{gmn}
\end{equation}
is the electron-phonon matrix element. Here, $\partial_{{\bf q}\nu}V$ is the derivative of the self-consistent electronic potential, and $\psi_{n{\bf k}}$ is the Bloch function for band $n$ and wave-vector ${\bf k}$.

For a single phonon mode $\nu$ with wave-vector ${\bf q}$, it is convenient to define a dimensionless electron-phonon coupling averaged over the Fermi-surface states with the density of states (DOS) $N(\varepsilon_F)=\sum_{n\bf k}\delta(\varepsilon_{n{\bf k}}-\varepsilon_F)$ as

\begin{figure}[!b]
	\centering
    \includegraphics[width=0.40\textwidth]{{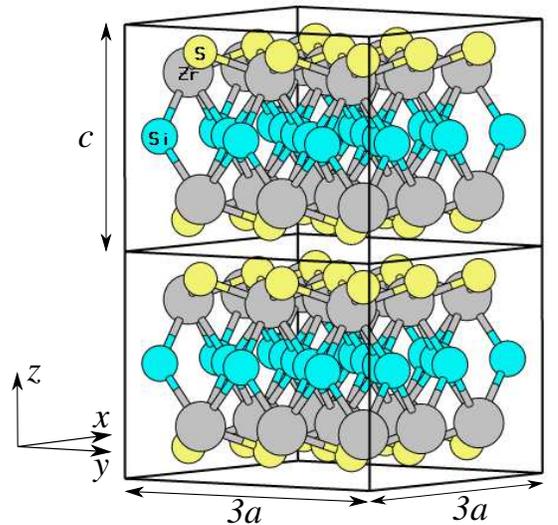}}
	\caption{Schematic of the ZrSiS crystal structure. Gray, cyan, and yellow balls correspond to Zr, Si, and S atoms, respectively. Depicted supercell contains 3$\times$3$\times$2=18 unit cells.}
	\label{fig1}
\end{figure}

\begin{equation}
\begin{split}
\lambda_{{\bf q}\nu} = 
\frac{1}{N(\varepsilon_F)\omega_{{\bf q}\nu}}\sum_{mn,{\bf k}} 
|g_{mn,\nu}({\bf k,q})|^2 \\ 
\times\delta(\varepsilon_{n{\bf k}}-\varepsilon_F)\delta(\varepsilon_{m{\bf k}+{\bf q}}-\varepsilon_F),
\end{split}
\label{lambda_q}
\end{equation}		
from which one can estimate the phonon linewidth as $\mathrm{Im}[\Pi_{{\bf q}{\nu}}] = \pi N(\varepsilon_F) \lambda_{{\bf q}\nu} \omega^2_{{\bf q}\nu}$.
From Eq.~(\ref{lambda_q}), one can also obtain a spectral representation of the electron-phonon interaction, known as the Eliashberg spectral function,
\begin{equation}
\alpha^{2}F( \omega) = \frac{1}{2} \sum_{{\bf q}\nu} 
\omega_{{\bf q}\nu} \lambda_{{\bf q}\nu}
\delta (\omega - \omega_{{\bf q}\nu}).
\label{a2f}
\end{equation}											

\begin{figure*}[!t]
	\centering
	\includegraphics[width=0.495\textwidth]{{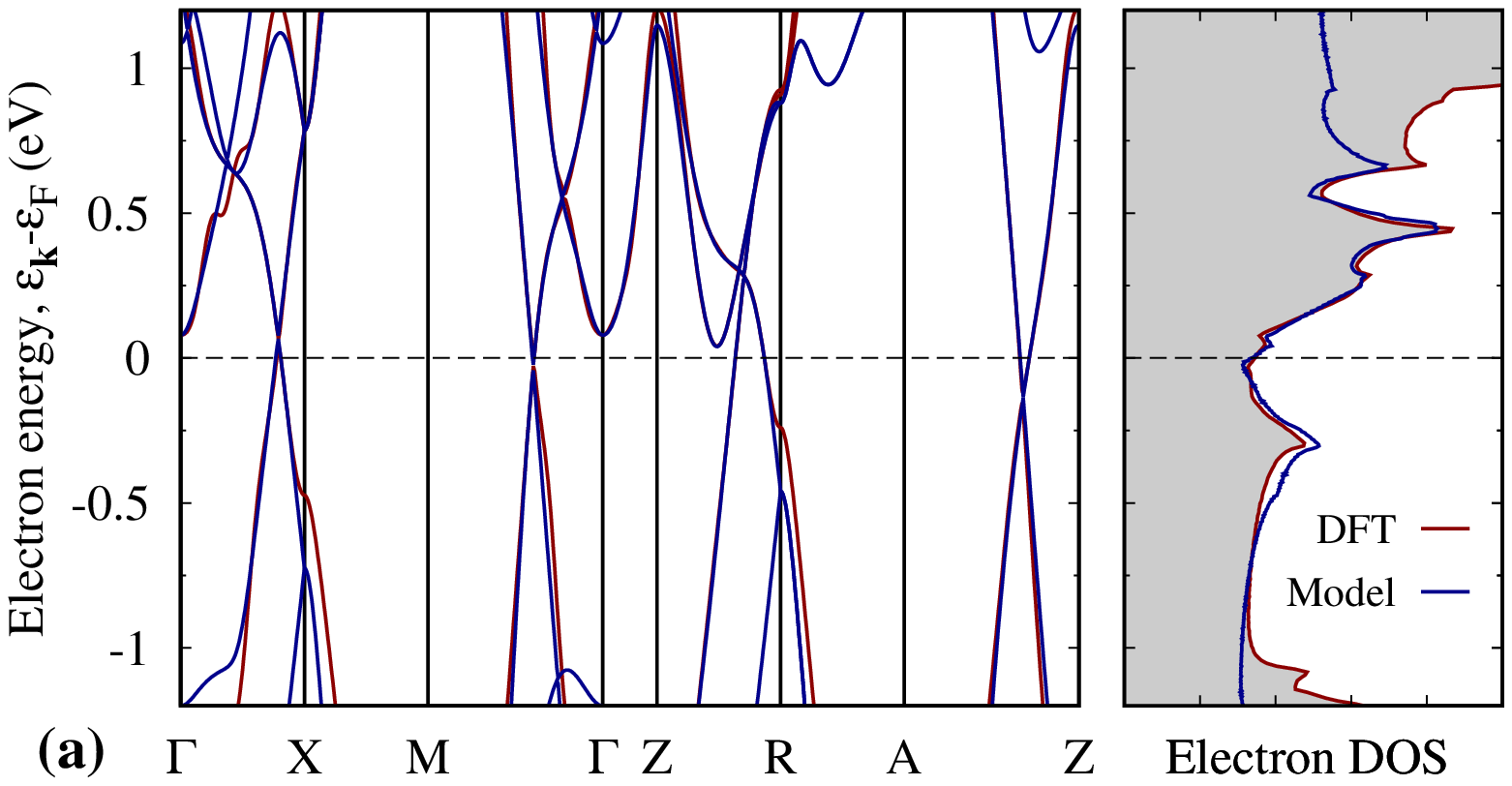}}
	\includegraphics[width=0.495\textwidth]{{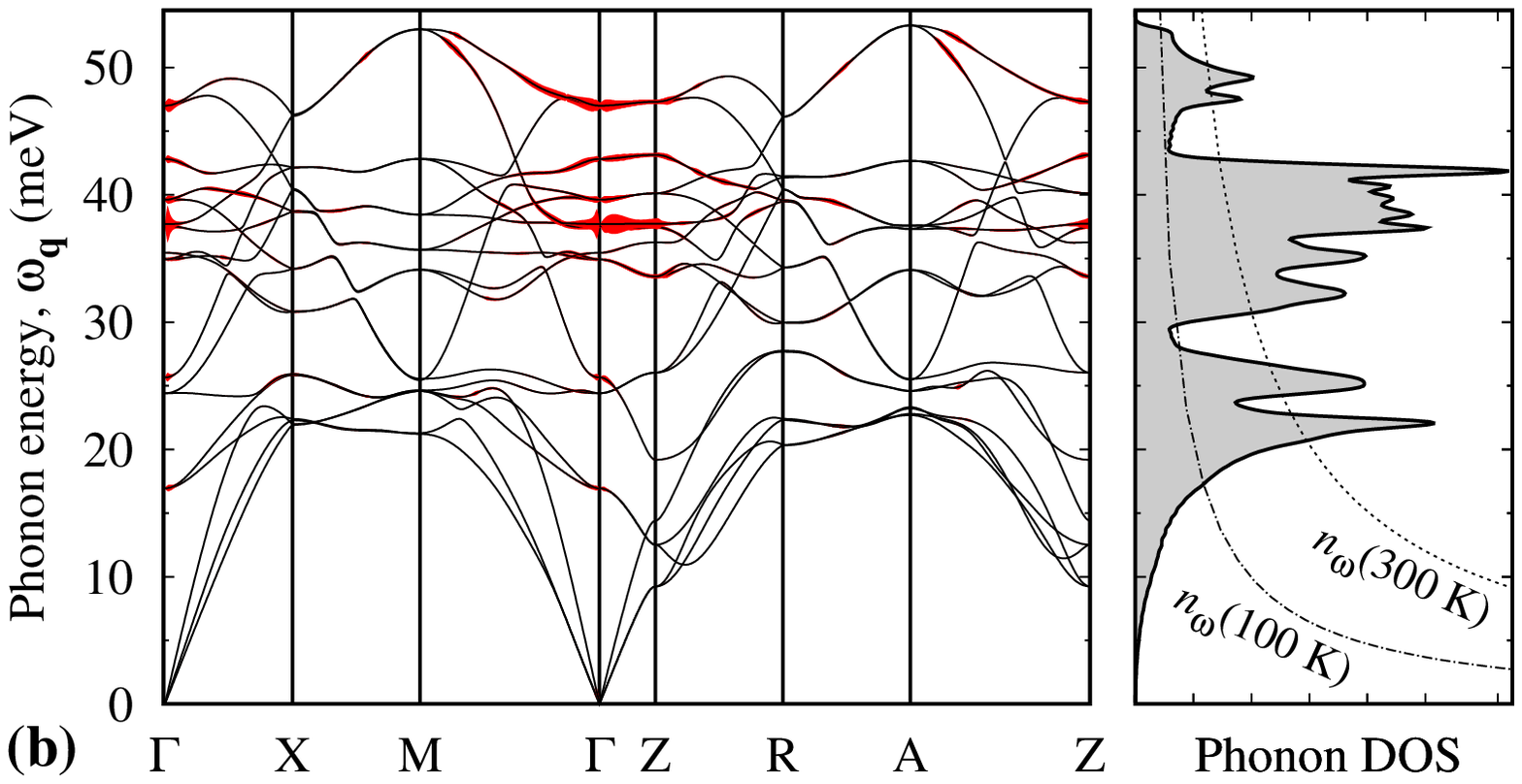}}
	\caption{(a) Band structure and electronic DOS of ZrSiS calculated using full DFT and a three-band TB model as described in the text. (b) Phonon dispersion and the corresponding DOS. A relative phonon linewidth $\mathrm{Im}[\Pi_{{\bf q}{\nu}}]$ 	is shown in the red color (not to scale).	The dashed lines are the equilibrium phonon distribution function $n_{\omega}(T)=(e^{\omega/T}-1)^{-1}$ shown for $T=100$ and 300 K.}
	\label{fig2}
\end{figure*}

Density-functional theory (DFT) electronic structure calculations and structural optimization were performed in this paper by means of the plane-wave {\sc quantum espresso} ({\sc qe}) code~\cite{Giannozzi2017}, using norm-conserving pseudopotentials \cite{gth,hgh}. Exchange  and correlation were treated within the local density approximation (LDA) \cite{lda-pz}. The kinetic energy cutoff for plane waves was set to 70 Ry, the Brillouin zone was sampled with a ($14\times14\times6$) Monkhorst-Pack {\bf k}-point mesh \cite{Monkhorst1976}, proved to be sufficient to achieve numerical accuracy. The crystal structure was fully relaxed with a threshold of \pten{1}{-12}~eV for total energies and \pten{1}{-12}~eV/\AA~ for forces. The electron-phonon matrix elements were calculated using density functional perturbation theory (DFPT) \cite{BaroniRMP} as implemented in {\sc qe}. The Brillouin zone for phonons within DFPT was sampled by a ($7\times7\times3$) {\bf q}-point mesh, and the self-consistency threshold of \pten{1}{-16}~eV was used. 
Spin-orbit coupling was not taken into account in the calculations, which is justified by its marginal influence on vibrational and electronic properties of ZrSiS at 
temperatures considered in this paper, which is demonstrated in the Supplemental Material (SM) \cite{SI}.
In order to achieve numerical accuracy of the Brillouin-zone integrals related to the electron-phonon coupling, we used an interpolation scheme based on the maximally localized Wannier functions \cite{Giustino2017,Marzari2012}. To this end, we focused on low-energy electronic states of ZrSiS and constructed a minimal tight-binding (TB) model described by a ($3\times3$) Hamiltonian matrix, which reproduces the electronic states in the vicinity of the Fermi energy with sufficient accuracy. The electronic self-energy was calculated on a ($144\times144\times32$) {\bf k}-point and ($32\times32\times12$) ${\bf q}$-point mesh, whereas different meshes were used to calculate the electron-phonon coupling strength and the Eliashberg function: ($384\times384\times32$) and ($24\times24\times8$), respectively. A smearing parameter of $\sigma=5$ meV was used for the Fermi surface averaging over the electron states. No artificial smearing effects are expected because the condition $\sigma<k_BT$ was fulfilled in all cases considered. The Eliashberg function was calculated using a phonon 
smearing of 0.05 meV.
The convergence behavior of the Brillouin-zone integrals with respect to the density of ${\bf k}$- and ${\bf q}$-point meshes is shown in the SM \cite{SI}.
All the computational parameters used in this paper were checked to be sufficient to reach numerical convergence.
The Wannier functions were constructed by means of the {\sc wannier90} code \cite{wannier90}, and the interpolation was performed within the {\sc epw} package \cite{Ponce2016}. The {\sc xcrysden} package was used to generate the crystal structure and the Fermi surface \cite{xcrysden}.

\section{Results and discussion}
\begin{figure}[tbp]
	\centering
    \includegraphics[width=0.47\textwidth]{{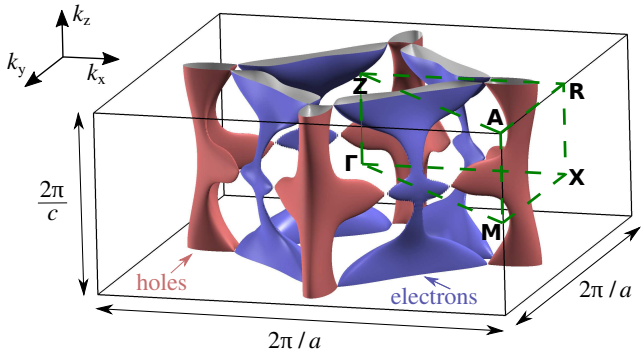}}
	\caption{Calculated Fermi surface of ZrSiS. The red and blue pockets denote valence (hole) and conduction (electron) states, respectively. The black solid lines mark boundaries of the Brillouin zone. The green dashed lines connect high-symmetry points.} 
	\label{fig_FS}
\end{figure}

{\label{sec3}}
ZrSiS is a layer crystal with tetragonal symmetry (point-group $D_{4h}$), schematically shown in Fig.~\ref{fig1}. Each layer of ZrSiS is composed of Si atoms sandwiched between Zr and S atoms, and the unit cell contains six atoms. The lattice parameters optimized within the LDA read $a=3.47$ and $c=7.92$ \AA.
In Fig.~\ref{fig2}(a), we show the band structure of ZrSiS and the corresponding electron DOS calculated using DFT-LDA and a three-band TB model constructed from the Wannier functions as described in Sec.~\ref{sec2}. 
A three-band model constitutes a minimal model for a reliable description of the electronic states in the vicinity of the Fermi energy. On the other hand, it appears as an optimal choice between the computational efficiency and the numerical accuracy within the Wannier interpolation scheme utilized in this paper.
The electronic structure of ZrSiS 
features linearly dispersing bands in the vicinity of the Fermi energy, crossing each other along multiple directions in the Brillouin zone, forming a set of continuous nodal lines (loops) in ${\bf k}$ space. Above the Fermi energy, there is a parabolic band at the $\Gamma$ point as well as along the $Z$--$R$ direction, which is expected to influence the charge carrier scattering in ZrSiS under electron doping and in the presence of tensile strain applied along the stacking direction \cite{Zhou}. 
The constructed TB model allows for a reliable description of the DFT bands in the energy range from $-$0.2 to +0.2 eV around the Fermi energy as well as their derivatives 
[see Fig.~\ref{fig5}(b) below].
Unlike graphene, ZrSiS exhibits a finite DOS at the Fermi energy, being a consequence of the continuous nodal line. Also, the position of the nodal line does not coincide with the Fermi energy, making the electron and hole states in ZrSiS not perfectly symmetric. The Fermi surface of ZrSiS is formed by four electron and four hole pockets, connected to each other at specific points on the $k_z=\pm \pi/c$ and $k_z=0$ plane as is shown in Fig.~\ref{fig_FS}.

\begin{figure}[b]
	\centering
	\mbox{
	\includegraphics[width=0.28\textwidth]{{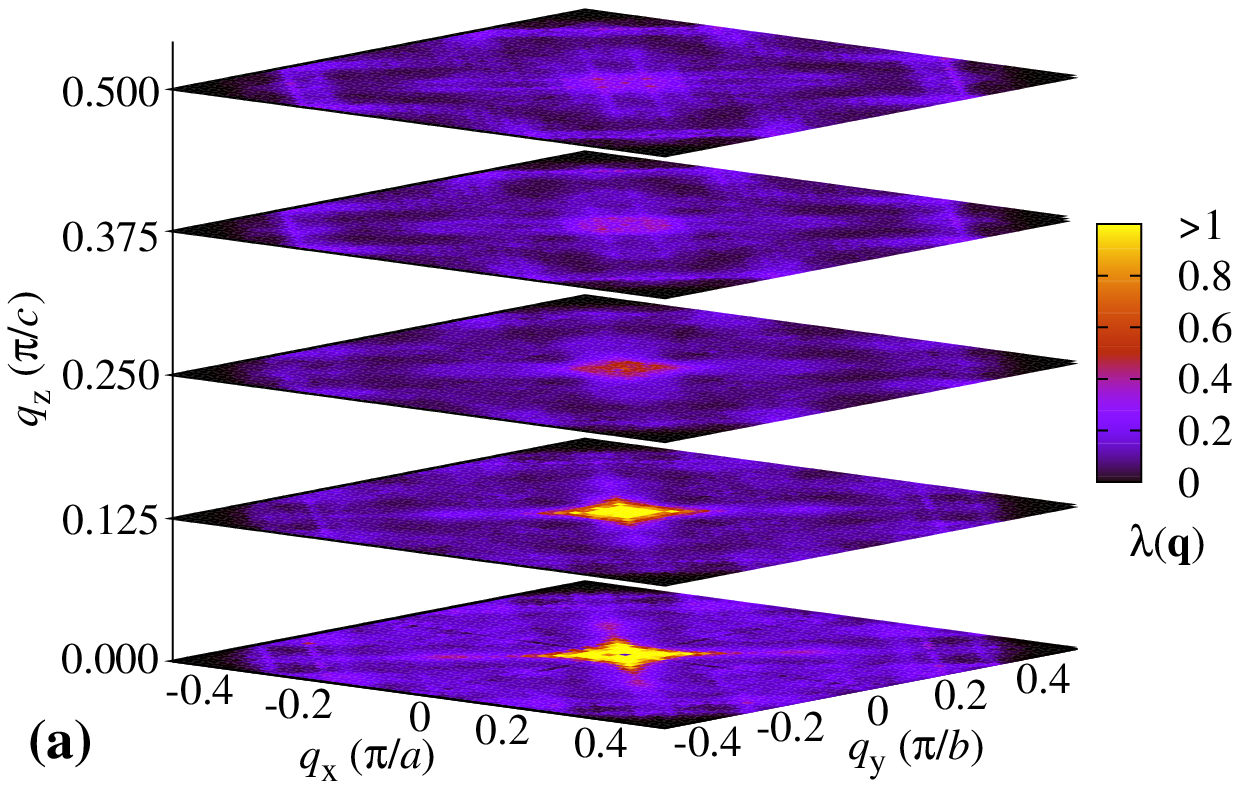}}
	\includegraphics[width=0.20\textwidth]{{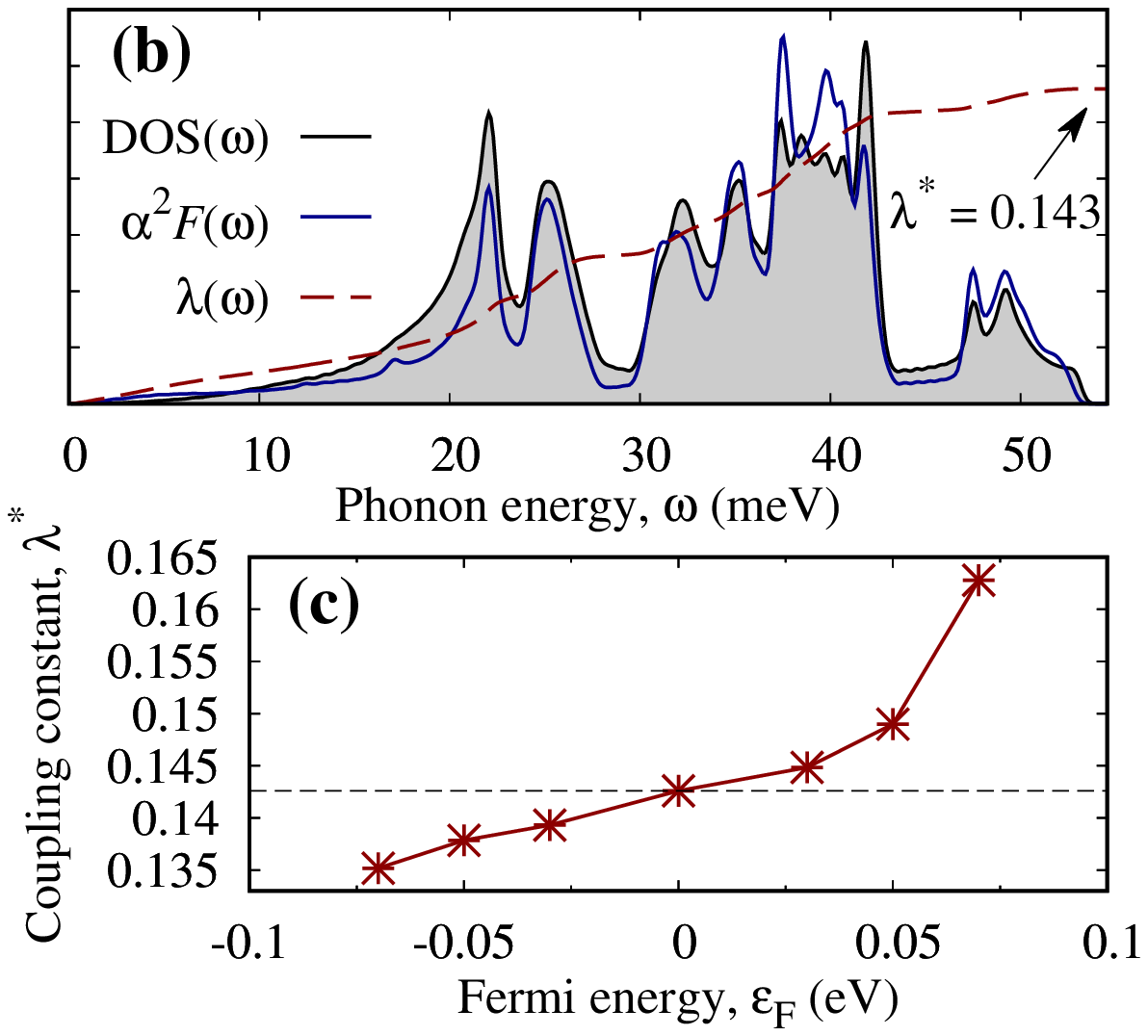}}
	}
	\caption{(a) Momentum-resolved electron-phonon coupling constant $\lambda_{\bf q}$. (b) Frequency-resolved electron-phonon spectral function $\alpha^2 F(\omega)$ and the integrated coupling constant $\lambda(\omega)=2\int d\omega ~ \omega^{-1}\alpha^2F(\omega)$ calculated for undoped ZrSiS ($\varepsilon_F=0$). (c) Total coupling constant $\lambda^* \equiv \lambda(\omega_{max})$ calculated as a function of the Fermi energy $\varepsilon_F$.}
	\label{fig3}
\end{figure}

\begin{figure}[tbp]
	\centering
	\includegraphics[width=0.50\textwidth]{{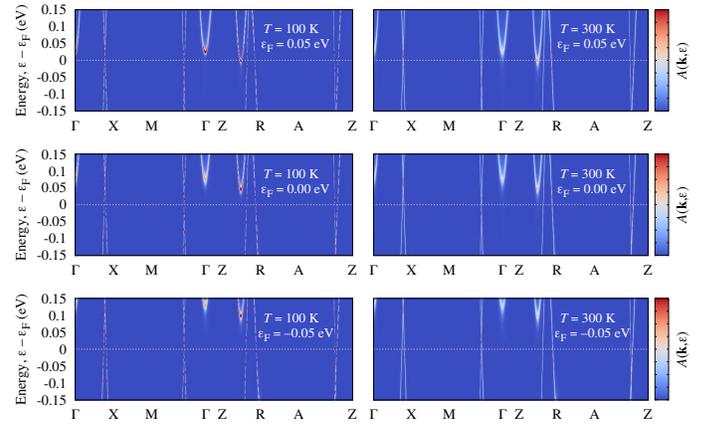}}
	\caption{Electron spectral function $A(\kv,\varepsilon)$ renormalized by the coupling with phonons in ZrSiS shown for different Fermi energies at $T=$ 100 and 300 K.}
	\label{fig4}
\end{figure}

The phonon dispersion and the corresponding DOS are shown in Fig.~\ref{fig2}(b), being in agreement with earlier studies \cite{Duman16,Zhou17,Singha18,Xue19}. One can see three acoustic branches with the linear dispersion $\omega_{\bf q}\sim |{\bf q}|$ and a series of optical branches ranging from 10 to 50 meV. Since the system is anisotropic, the dispersion of acoustic modes is strongly dependent on phonon polarization. For phonons propagating in the [001] direction ($z$), the sound velocity $v^z_s=\partial \omega({\bf q})/\partial q_z |_{{\bf q}=0}$
is estimated to be 6.8 and 4.3 km/s for out-of-plane and in-plane polarizations, respectively. Along the [100] propagating direction, these values are roughly 10\% higher.
The optical phonons in ZrSiS are characterized by a number of flat branches, which give rise to several peaks (van Hove singularities) in phonon DOS.
In the energy region between 20 and 25 meV, there are two sharp peaks originating from low-energy optical branches associated mainly with Zr vibrations. 
These branches are expected to provide a significant contribution to phonon population at room temperature as can be deduced from the phonon distribution function depicted along with DOS in Fig.~\ref{fig2}(b).
In the same figure, we also show a broadening of phonon lines shown together with the noninteracting dispersion. The maximum linewidth is observed for 
optical phonons with small wave vectors but with energies $\omega_{{{\bf q}\nu}} > 30$ meV, indicating a relatively small population of interacting states at room temperature and below.

We now turn to the electron-phonon coupling in ZrSiS. In Fig.~\ref{fig3}(a), we show the electron-phonon coupling strength $\lambda_{\bf q}$ resolved over the Brillouin zone. One can see that the main contribution corresponds to the coupling with long-wavelength phonons, maximizing $\lambda_{\bf q}$ around the $\Gamma$ point. This behavior can be attributed to the structure of the Fermi surface, which exhibits elongated electron and hole pockets (Fig.~\ref{fig_FS}), favoring transitions with small momentum transfer. Although
one may expect strong nesting effects from the diamond-shaped Fermi surface \cite{Ali16}, these effects are not pronounced in the electron-phonon coupling, indicating small interaction weight at the corresponding momentum transfers. Figure \ref{fig3}(b) shows the 
Eliashberg function of ZrSiS, which closely follows phonon DOS, suggesting weak electron-phonon coupling and the absence of any significant many-body renormalization. Indeed, the integrated electron-phonon coupling constant $\lambda^*=\sum_{{\bf q}}\lambda_{\bf q}\approx 0.14$ is small, leading to a negligible mass enhancement $m^*/m=1+\lambda^*$. The coupling constant is weakly dependent on the Fermi energy as is shown in Fig.~\ref{fig3}(c). It is worth noting that similar small values for $\lambda$ have been recently reported in
Ref.~\cite{Xue19} yet obtained from a simplified model taking only one optical phonon mode into account focusing on the surface states of ZrSiS.
 
Weak electron-phonon coupling is favorable in terms of charge carrier transport as it generally ensures a low scattering rate. On the other hand, weak coupling suppresses phonon-mediated electron pairing, playing a key role in conventional superconductivity. The calculated 
electron-phonon coupling constant $\lambda$ allows one to make a rough estimate of the superconducting transition temperature within the Allen and Dynes formula \cite{AllenDynes}. Based on this estimation, we conclude that phonon-mediated superconductivity is unlikely
in ZrSiS even on the microkelvin temperature scale. This finding is consistent with the experiments, observing no superconductivity down to 2 K even at high pressure up to 20 GPa \cite{VanGennep19}. We note that this does not exclude other pairing mechanisms (e.g., excitonic 
pairing), which may become relevant at low enough temperatures \cite{Scherer18}.

Let us now discuss the influence of the electron-phonon interaction on the electronic structure.
Figure \ref{fig4} shows doping dependence of the electron spectral function calculated in the presence of electron-phonon coupling in ZrSiS for two characteristic temperatures: $T=100$ and 300 K. As is expected from small electron-phonon coupling strength, 
the spectral function closely resembles noninteracting bands shown in Fig.~\ref{fig2}. Apart from the finite linewidth, one can see prominent kinks in the band structure, clearly distinguishable at $T=100$ K. Rigid-shift doping does not lead to any
significant modification of the spectral function, despite some variation of the electron-phonon coupling strength [Fig.~\ref{fig3}(c)]. In the relevant energy region, the linewidth is essentially independent of {\bf k}, showing no anomalies near the nodal lines.
This indicates that electron-phonon scattering in ZrSiS is similar along different directions. In the vicinity of the Fermi energy, the imaginary part of the self-energy $\mathrm{Im}[\Sigma_{n{\bf k}}(\omega,T)]$ at $T=300$~K varies between 10 and 25 meV depending on the 
${\bf k}$ point and band index $n$, whereas, at $T=100$~K the variation reduces to 0.7--1.5 meV.

\begin{figure}[tbp]
	\centering
	\includegraphics[width=0.50\textwidth]{{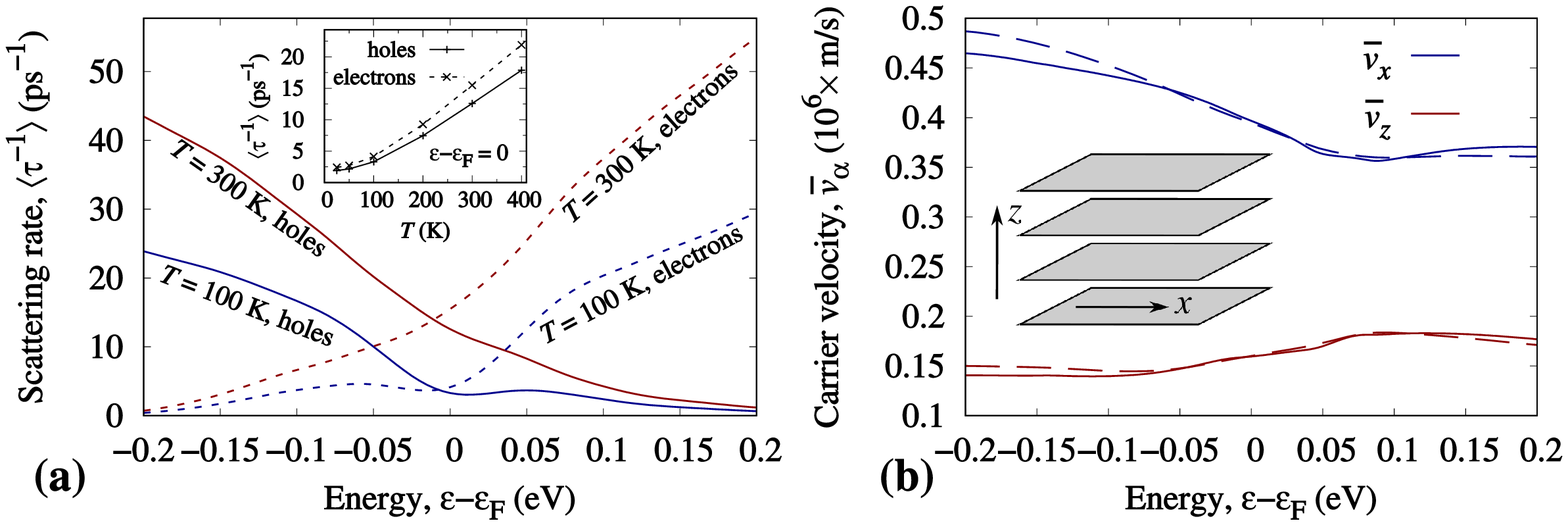}}
	\caption{(a) Energy dependence of the averaged scattering rate calculated for holes and electrons in charge neutral ZrSiS ($\varepsilon_F=0$) at $T=100$ and 300 K. The inset shows temperature dependence of the scattering rate for $\varepsilon=\varepsilon_F$. (b) Energy-resolved $x$ and $z$ components of the carrier velocity calculated as $\bar{v}_{\alpha}=\sqrt{\langle v^2_{\alpha} \rangle}$, where $\langle ... \rangle$ denotes the Fermi-surface average. The dashed lines in (b) show the result obtained using full DFT Hamiltonian, i.e., without constructing a TB model.}
	\label{fig5}
\end{figure}

The electronic self-energy resolved over the Brillouin zone allows us to restore the energy dependence of the electron and hole scattering rates. To this end, we average $\tau^{-1}_{n{\bf k}}$ given by Eq.~(\ref{tau}) over the states with a constant energy as 
\begin{equation}
	\langle \tau^{-1} \rangle = \frac{1}{N(\varepsilon)} \sum_{n\bf k}\tau^{-1}_{n\bf k}\delta(\varepsilon_{n\bf k}-\varepsilon),
	\label{average_tau}
\end{equation}
where $N(\varepsilon)$ is the electron DOS at energy $\varepsilon$. We note that, generally speaking, $\varepsilon$ in Eq.~(\ref{average_tau}) does not have the meaning of the Fermi energy because $\tau^{-1}_{n\bf k}$ being a function of ${\bf k}$ is determined by the Fermi 
energy [see Fig.~\ref{fig4}]. In Fig.~\ref{fig5}(a), we show the averaged scattering rate calculated separately for holes and electrons in ZrSiS as a function of energy for $T=100$ and 300 K. For the temperatures considered, we observe similar energy dependence
with the main difference being the magnitude of the scattering rate. The hole and electron curves are almost symmetric and cross near the zero energy. Thus, electrons and holes provide nearly equal contribution to the scattering rate, which is not surprising because both
states have comparable weight at the Fermi surface [see Fig.~\ref{fig_FS}]. It is interesting to note that the electron-hole symmetry in ZrSiS can be broken by uniaxial strain applied along the [001] direction \cite{Zhou}. The energy dependence of the scattering rate exhibits 
a monotonic behavior, which is almost linear at $T=300$ K but shows small deviations in the vicinity of the zero energy at $T=100$ K. Overall, the energy dependence resembles electron DOS, which is not unusual for the case of weak electron-phonon coupling.
Quantitatively, the room-temperature scattering rate in ZrSiS is comparable with that in typical noble metals (e.g., copper) \cite{Louie16}.

In the inset in Fig.~\ref{fig5}, one can see temperature dependence of the scattering rate shown for $\varepsilon=\varepsilon_F$.
The dependence mainly stems from the phonon occupation factor $n_{\bf q \nu}(T)$ entering the expression for self-energy [Eq.~(\ref{selfenergy})]. In the temperature range considered, phonons can be considered classically, meaning $n_{\qv\nu} \simeq k_BT/\hbar\omega_{\qv\nu}$ and
ensuring linear dependence of the electron linewidth $\mathrm{Im}[\Sigma_{{\bf q}\nu}(T)]$ with temperature. At $T\lesssim100$ K one can see deviations from the linear dependence related to the low-$T$ behavior of the phonon distribution. The obtained dependence is in 
agreement with experiments on zero-field resistivity in ZrSiS measured at various temperatures (e.g., see Ref.~\cite{Singha17}). 
Here, we do not explicitly consider low-$T$ range mainly because the Boltzmann transport theory is limited in this regime. Also, other scattering mechanisms become dominant at low $T$, such as electron-electron and electron-impurity scattering. 
Nevertheless, it is interesting to extrapolate the data to the zero-temperature limit after which we obtain a finite scattering rate. 
This can be understood as a residual scattering on zero-point vibrations, allowed by the approximation for self-energy [Eq.~(\ref{selfenergy})] used in this paper.
Our estimation of the corresponding relaxation time yields $\tau_0 \approx 0.25$ ps, which is an order of magnitude larger than the value obtained from the Hall measurements in Ref.~\cite{Novak19}. Expectedly, zero-temperature scattering is likely to be governed by 
other mechanisms.
Apart from obvious scattering sources, such as impurities or defects, electron-electron interaction could be especially important in ZrSiS at low $T$ because it may cause considerable renormalization of the electron energy spectrum \cite{Pezzini18,Rudenko18}. 
We also note that, under realistic experimental conditions, surface states might provide a nonzero contribution to the scattering, especially in thin samples, as suggested by recent reports on surface chemistry in ZrSiS \cite{Boukhvalov19}.

\begin{figure}[t]
	\centering
	\includegraphics[width=0.5\textwidth]{{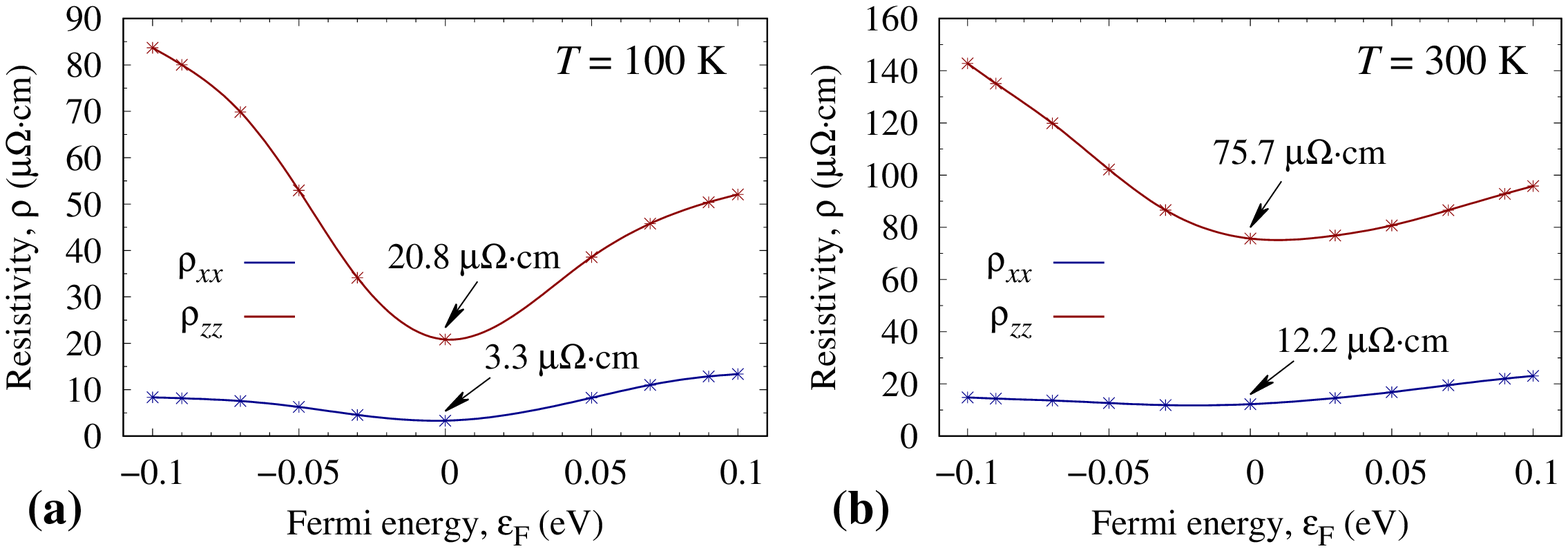}}
	\caption{In-plane ($\rho_{xx}$) and out-of-plane ($\rho_{zz}$) components of the resistivity tensor in ZrSiS shown as a function of the Fermi energy for (a) $T=100$ K and (b) $T=300$ K.}
	\label{fig6}
\end{figure}

Apart from the scattering rate, transport properties are largely determined by the carrier velocities. In Fig.~\ref{fig5}(b), we show energy dependence of the averaged carrier velocities in ZrSiS calculated along the in-plane [100] and out-of-plane [001] 
directions of the crystal.
In both cases, the energy dependence is weak, and the anisotropy $\bar{v}_{x} / \bar{v}_{z}$ varies between two and three depending of the energy. Along the in-plane direction, the obtained values are around 0.4$\times$10$^{6}$ m/s, which is somewhat smaller than the Fermi velocity
in graphene. It should be noted that the DFT values for the carrier velocity may be underestimated. In graphene, many-body corrections applied at the $GW$ level of theory lead to an $\sim$17\% enhancement of the Fermi velocity \cite{Trevi08}.
Similar magnitude of renormalization has recently been obtained for a similar nodal-line semimetal ZrSiSe \cite{Shao19}. Since the conductivity is quadratically dependent on the carrier velocity [Eq.~(\ref{sigma})], we expect that the resistivity calculated in our
paper is likely overestimated by up to 30\% compared to $GW$ calculations. It is worth noting, however, that this expectation is not consistent with experimental observations, suggesting a suppression of the carrier velocities in ZrSiS \cite{Pezzini18} and 
ZrSiSe \cite{Shao19}, at least, in the low-$T$ regime.

Finally, we consider resistivities and their dependence of the rigid shift doping (Fermi energy). In Fig.~\ref{fig6}, one can see the resistivities $\rho=1/\sigma$ calculated along the [100] and [001] directions in ZrSiS. The in-plane resistivity
$\rho_{xx}$ is weakly dependent on the Fermi energy with the minimum values being 3.3 and 12.2 $\mu \Omega$\,cm at $T=100$ and 300 K, respectively. In contrast, the out-of-plane resistivity $\rho_{zz}$ exhibits a more pronounced energy dependence, varying by more than
factor of two in the considered energy range. At the neutrality point, out-of-plane resistivity values are significantly higher, reaching 20.8 and 75.7 $\mu \Omega$\,cm for $T=100$ and 300 K, respectively. The corresponding anisotropy ratio is $\rho_{zz}/\rho_{xx}\sim 6.3$--$8.8$ 
depending on the energy and temperature considered. Temperature dependence of the resistivity can be easily restored from the inset in Fig.~\ref{fig5}(a) keeping in mind that
within the approach utilized in this paper $T$ dependence originates predominantly from the scattering rate, $\rho(T) \sim \langle \tau^{-1}(T) \rangle$. As is expected for conventional
conductors, at moderate temperatures, we obtain linear dependence $\rho \sim T$, in accordance with experiments \cite{Ali16,Singha17,Sankar17,Hu17,VanGennep19,Shirer19}.

It is instructive to compare the calculated resistivity with experimental data, which are broadly presented in the literature. The measurements of zero-field resistivity are mostly limited to the in-plane direction, demonstrating a rather large spread of values.
Most of the experimental studies report in-plane resistivity in ZrSiS at room temperature to be around $15\pm1$ $\mu \Omega$\,cm \cite{Ali16,Singha17,Sankar17,Shirer19}. 
These values are in good agreement with our calculations (12.2 $\mu \Omega$\,cm), 
keeping in mind that only electron-phonon scattering mechanism is taken into account. 
Some studies, however, report considerably lower values: 8 \cite{Novak19} and 10 $\mu \Omega$\,cm \cite{VanGennep19}, which appears somewhat unexpecting taking into 
account that the values calculated in our paper should be considered as a lower bound for resistivity because other scattering mechanisms are present in experiments, increasing the 
resistivity. 
Apart from experimental uncertainty, the calculated resistivity could
be overestimated due to ignoring of many-body effects, which are expected to renormalize the carrier velocities and lead to lower resistivities as we already discussed above.
In contrast, Ref.~\cite{Hu17} reports much higher resistivity in ZrSiS with $\rho_{xx}\approx 27$ $\mu \Omega$\,cm, which is apparently in conflict with other available measurements, and
probably points to a poor sample quality. Accurate measurements along the [001] direction have proven to be way more challenging. In a recent study by Shirer \emph{et al.} \cite{Shirer19}, magnetotransport experiments have been performed on ZrSiS microstructures with well-defined 
geometry and high sample quality \cite{Shirer19}, yielding $\rho_{zz}\approx 110$ $\mu \Omega$\,cm at $T=300$ K and $\rho_{zz}/\rho_{xx}\approx8$ showing weak dependence of the temperature. Both values are in good agreement with the values obtained in this paper, 
taking into account the reservations made above. It is fair to say that a significantly larger resistivity anisotropy ($\sim$50) has been obtained by Novak \emph{et al.} in Ref.~\cite{Novak19}.
However, by far larger residual resistivity observed along the [001] direction compared to Ref.~\cite{Shirer19} makes us believe that those findings are less reliable as far as pristine ZrSiS is concerned. 
Giant anisotropy could point to possible contamination between the layers, typical to van der Waals heterostructures \cite{Rooney}, resulting in an enhanced electron-impurity scattering along the [001] direction.

\section{Conclusion}{\label{sec4}}
To conclude, we have performed a first-principles study of electron-phonon interactions and related transport properties in nodal-line semimetal ZrSiS.
To this end, we used density functional perturbation theory in combination with the semiclassical Boltzmann transport formalism. We found that
electron-phonon interaction in ZrSiS is weak with the coupling constant $\lambda\sim$~0.1 being virtually independent of external doping. The main contribution
to the interaction originates from long-wavelength optical phonons, mainly associated with Zr vibrations. The electron spectral function calculated
within the Migdal approximation shows no significant renormalization in the presence of electron-phonon interaction and is only characterized by
a finite linewidth compared to the noninteracting spectrum. Although the scattering rate is found to be essentially independent of the direction of electron propagation,
the carrier velocities are strongly anisotropic between the in-plane [001] and out-of-plane [001] directions. This gives rise to a pronounced anisotropy in the transport
properties with the resistivity ratio around $\rho_{zz}/\rho_{xx}\sim 7.5$. Quantitatively, the phonon-limited zero-field resistivity along the in-plane direction is estimated
to be $\rho_{xx}\approx12$ $\mu\Omega$\,cm at room temperature, which has a comparable contribution from both hole and electron carrier channels. The relatively low value can be directly attributed to a combination of
weak electron-phonon coupling and high carrier velocities.
Overall, the obtained values are found to be in good agreement with available experimental data, although a notable discrepancy between the experimental results of different research groups should be mentioned.
As in conventional metallic conductors, low resistivity in ZrSiS excludes electron-phonon coupling from being a mechanism for superconductivity at any realistic temperatures.

Our paper constitutes a step forward toward microscopic understanding of charge carrier scattering and transport properties in ZrSiS, and may be useful for further theoretical and experimental
studies of electronic transport properties of this material and its analogs.

\FloatBarrier
\begin{acknowledgments}
A.N.R. acknowledges computational resources at the Radboud University partially funded by FLAG-ERA JTC2017 Project GRANSPORT.
\end{acknowledgments}

\bibliographystyle{apsrev4-1}
\bibliography{main}

\begin{thebibliography}{58}%
\makeatletter
\providecommand \@ifxundefined [1]{%
 \@ifx{#1\undefined}
}%
\providecommand \@ifnum [1]{%
 \ifnum #1\expandafter \@firstoftwo
 \else \expandafter \@secondoftwo
 \fi
}%
\providecommand \@ifx [1]{%
 \ifx #1\expandafter \@firstoftwo
 \else \expandafter \@secondoftwo
 \fi
}%
\providecommand \natexlab [1]{#1}%
\providecommand \enquote  [1]{``#1''}%
\providecommand \bibnamefont  [1]{#1}%
\providecommand \bibfnamefont [1]{#1}%
\providecommand \citenamefont [1]{#1}%
\providecommand \href@noop [0]{\@secondoftwo}%
\providecommand \href [0]{\begingroup \@sanitize@url \@href}%
\providecommand \@href[1]{\@@startlink{#1}\@@href}%
\providecommand \@@href[1]{\endgroup#1\@@endlink}%
\providecommand \@sanitize@url [0]{\catcode `\\12\catcode `\$12\catcode
  `\&12\catcode `\#12\catcode `\^12\catcode `\_12\catcode `\%12\relax}%
\providecommand \@@startlink[1]{}%
\providecommand \@@endlink[0]{}%
\providecommand \url  [0]{\begingroup\@sanitize@url \@url }%
\providecommand \@url [1]{\endgroup\@href {#1}{\urlprefix }}%
\providecommand \urlprefix  [0]{URL }%
\providecommand \Eprint [0]{\href }%
\providecommand \doibase [0]{http://dx.doi.org/}%
\providecommand \selectlanguage [0]{\@gobble}%
\providecommand \bibinfo  [0]{\@secondoftwo}%
\providecommand \bibfield  [0]{\@secondoftwo}%
\providecommand \translation [1]{[#1]}%
\providecommand \BibitemOpen [0]{}%
\providecommand \bibitemStop [0]{}%
\providecommand \bibitemNoStop [0]{.\EOS\space}%
\providecommand \EOS [0]{\spacefactor3000\relax}%
\providecommand \BibitemShut  [1]{\csname bibitem#1\endcsname}%
\let\auto@bib@innerbib\@empty
\bibitem [{\citenamefont {Young}\ \emph {et~al.}(2012)\citenamefont {Young},
  \citenamefont {Zaheer}, \citenamefont {Teo}, \citenamefont {Kane},
  \citenamefont {Mele},\ and\ \citenamefont {Rappe}}]{Young12}%
  \BibitemOpen
  \bibfield  {author} {\bibinfo {author} {\bibfnamefont {S.~M.}\ \bibnamefont
  {Young}}, \bibinfo {author} {\bibfnamefont {S.}~\bibnamefont {Zaheer}},
  \bibinfo {author} {\bibfnamefont {J.~C.~Y.}\ \bibnamefont {Teo}}, \bibinfo
  {author} {\bibfnamefont {C.~L.}\ \bibnamefont {Kane}}, \bibinfo {author}
  {\bibfnamefont {E.~J.}\ \bibnamefont {Mele}}, \ and\ \bibinfo {author}
  {\bibfnamefont {A.~M.}\ \bibnamefont {Rappe}},\ }\bibfield  {booktitle}
  {\emph {\bibinfo {booktitle} {Dirac Semimetal in Three Dimensions}},\ }\href
  {\doibase 10.1103/PhysRevLett.108.140405} {\bibfield  {journal} {\bibinfo
  {journal} {Phys. Rev. Lett.}\ }\textbf {\bibinfo {volume} {108}},\ \bibinfo
  {pages} {140405} (\bibinfo {year} {2012})}\BibitemShut {NoStop}%
\bibitem [{\citenamefont {Liu}\ \emph {et~al.}(2014)\citenamefont {Liu},
  \citenamefont {Zhou}, \citenamefont {Zhang}, \citenamefont {Wang},
  \citenamefont {Weng}, \citenamefont {Prabhakaran}, \citenamefont {Mo},
  \citenamefont {Shen}, \citenamefont {Fang}, \citenamefont {Dai},
  \citenamefont {Hussain},\ and\ \citenamefont {Chen}}]{Liu14}%
  \BibitemOpen
  \bibfield  {author} {\bibinfo {author} {\bibfnamefont {Z.~K.}\ \bibnamefont
  {Liu}}, \bibinfo {author} {\bibfnamefont {B.}~\bibnamefont {Zhou}}, \bibinfo
  {author} {\bibfnamefont {Y.}~\bibnamefont {Zhang}}, \bibinfo {author}
  {\bibfnamefont {Z.~J.}\ \bibnamefont {Wang}}, \bibinfo {author}
  {\bibfnamefont {H.~M.}\ \bibnamefont {Weng}}, \bibinfo {author}
  {\bibfnamefont {D.}~\bibnamefont {Prabhakaran}}, \bibinfo {author}
  {\bibfnamefont {S.-K.}\ \bibnamefont {Mo}}, \bibinfo {author} {\bibfnamefont
  {Z.~X.}\ \bibnamefont {Shen}}, \bibinfo {author} {\bibfnamefont
  {Z.}~\bibnamefont {Fang}}, \bibinfo {author} {\bibfnamefont {X.}~\bibnamefont
  {Dai}}, \bibinfo {author} {\bibfnamefont {Z.}~\bibnamefont {Hussain}}, \ and\
  \bibinfo {author} {\bibfnamefont {Y.~L.}\ \bibnamefont {Chen}},\ }\bibfield
  {booktitle} {\emph {\bibinfo {booktitle} {{Discovery of a Three-Dimensional
  Topological Dirac Semimetal, Na$_3$Bi}}},\ }\href {\doibase
  10.1126/science.1245085} {\bibfield  {journal} {\bibinfo  {journal}
  {Science}\ }\textbf {\bibinfo {volume} {343}},\ \bibinfo {pages} {864}
  (\bibinfo {year} {2014})}\BibitemShut {NoStop}%
\bibitem [{\citenamefont {Borisenko}\ \emph {et~al.}(2014)\citenamefont
  {Borisenko}, \citenamefont {Gibson}, \citenamefont {Evtushinsky},
  \citenamefont {Zabolotnyy}, \citenamefont {B\"uchner},\ and\ \citenamefont
  {Cava}}]{Borisenko14}%
  \BibitemOpen
  \bibfield  {author} {\bibinfo {author} {\bibfnamefont {S.}~\bibnamefont
  {Borisenko}}, \bibinfo {author} {\bibfnamefont {Q.}~\bibnamefont {Gibson}},
  \bibinfo {author} {\bibfnamefont {D.}~\bibnamefont {Evtushinsky}}, \bibinfo
  {author} {\bibfnamefont {V.}~\bibnamefont {Zabolotnyy}}, \bibinfo {author}
  {\bibfnamefont {B.}~\bibnamefont {B\"uchner}}, \ and\ \bibinfo {author}
  {\bibfnamefont {R.~J.}\ \bibnamefont {Cava}},\ }\bibfield  {booktitle} {\emph
  {\bibinfo {booktitle} {Experimental Realization of a Three-Dimensional Dirac
  Semimetal}},\ }\href {\doibase 10.1103/PhysRevLett.113.027603} {\bibfield
  {journal} {\bibinfo  {journal} {Phys. Rev. Lett.}\ }\textbf {\bibinfo
  {volume} {113}},\ \bibinfo {pages} {027603} (\bibinfo {year}
  {2014})}\BibitemShut {NoStop}%
\bibitem [{\citenamefont {Xu}\ \emph {et~al.}(2015)\citenamefont {Xu},
  \citenamefont {Belopolski}, \citenamefont {Alidoust}, \citenamefont
  {Neupane}, \citenamefont {Bian}, \citenamefont {Zhang}, \citenamefont
  {Sankar}, \citenamefont {Chang}, \citenamefont {Yuan}, \citenamefont {Lee},
  \citenamefont {Huang}, \citenamefont {Zheng}, \citenamefont {Ma},
  \citenamefont {Sanchez}, \citenamefont {Wang}, \citenamefont {Bansil},
  \citenamefont {Chou}, \citenamefont {Shibayev}, \citenamefont {Lin},
  \citenamefont {Jia},\ and\ \citenamefont {Hasan}}]{Xu15}%
  \BibitemOpen
  \bibfield  {author} {\bibinfo {author} {\bibfnamefont {S.-Y.}\ \bibnamefont
  {Xu}}, \bibinfo {author} {\bibfnamefont {I.}~\bibnamefont {Belopolski}},
  \bibinfo {author} {\bibfnamefont {N.}~\bibnamefont {Alidoust}}, \bibinfo
  {author} {\bibfnamefont {M.}~\bibnamefont {Neupane}}, \bibinfo {author}
  {\bibfnamefont {G.}~\bibnamefont {Bian}}, \bibinfo {author} {\bibfnamefont
  {C.}~\bibnamefont {Zhang}}, \bibinfo {author} {\bibfnamefont
  {R.}~\bibnamefont {Sankar}}, \bibinfo {author} {\bibfnamefont
  {G.}~\bibnamefont {Chang}}, \bibinfo {author} {\bibfnamefont
  {Z.}~\bibnamefont {Yuan}}, \bibinfo {author} {\bibfnamefont {C.-C.}\
  \bibnamefont {Lee}}, \bibinfo {author} {\bibfnamefont {S.-M.}\ \bibnamefont
  {Huang}}, \bibinfo {author} {\bibfnamefont {H.}~\bibnamefont {Zheng}},
  \bibinfo {author} {\bibfnamefont {J.}~\bibnamefont {Ma}}, \bibinfo {author}
  {\bibfnamefont {D.~S.}\ \bibnamefont {Sanchez}}, \bibinfo {author}
  {\bibfnamefont {B.}~\bibnamefont {Wang}}, \bibinfo {author} {\bibfnamefont
  {A.}~\bibnamefont {Bansil}}, \bibinfo {author} {\bibfnamefont
  {F.}~\bibnamefont {Chou}}, \bibinfo {author} {\bibfnamefont {P.~P.}\
  \bibnamefont {Shibayev}}, \bibinfo {author} {\bibfnamefont {H.}~\bibnamefont
  {Lin}}, \bibinfo {author} {\bibfnamefont {S.}~\bibnamefont {Jia}}, \ and\
  \bibinfo {author} {\bibfnamefont {M.~Z.}\ \bibnamefont {Hasan}},\ }\bibfield
  {booktitle} {\emph {\bibinfo {booktitle} {Discovery of a Weyl fermion
  semimetal and topological Fermi arcs}},\ }\href {\doibase
  10.1126/science.aaa9297} {\bibfield  {journal} {\bibinfo  {journal}
  {Science}\ }\textbf {\bibinfo {volume} {349}},\ \bibinfo {pages} {613}
  (\bibinfo {year} {2015})}\BibitemShut {NoStop}%
\bibitem [{\citenamefont {Castro~Neto}\ \emph {et~al.}(2009)\citenamefont
  {Castro~Neto}, \citenamefont {Guinea}, \citenamefont {Peres}, \citenamefont
  {Novoselov},\ and\ \citenamefont {Geim}}]{Castro-Neto09}%
  \BibitemOpen
  \bibfield  {author} {\bibinfo {author} {\bibfnamefont {A.~H.}\ \bibnamefont
  {Castro~Neto}}, \bibinfo {author} {\bibfnamefont {F.}~\bibnamefont {Guinea}},
  \bibinfo {author} {\bibfnamefont {N.~M.~R.}\ \bibnamefont {Peres}}, \bibinfo
  {author} {\bibfnamefont {K.~S.}\ \bibnamefont {Novoselov}}, \ and\ \bibinfo
  {author} {\bibfnamefont {A.~K.}\ \bibnamefont {Geim}},\ }\bibfield
  {booktitle} {\emph {\bibinfo {booktitle} {The electronic properties of
  graphene}},\ }\href {\doibase 10.1103/RevModPhys.81.109} {\bibfield
  {journal} {\bibinfo  {journal} {Rev. Mod. Phys.}\ }\textbf {\bibinfo {volume}
  {81}},\ \bibinfo {pages} {109} (\bibinfo {year} {2009})}\BibitemShut
  {NoStop}%
\bibitem [{\citenamefont {Katsnelson}(2012)}]{Katsnelson-Book}%
  \BibitemOpen
  \bibfield  {author} {\bibinfo {author} {\bibfnamefont {M.~I.}\ \bibnamefont
  {Katsnelson}},\ }\href@noop {} {\emph {\bibinfo {title} {Graphene: Carbon in
  Two Dimensions}}}\ (\bibinfo  {publisher} {Cambridge University Press,
  Cambridge, U.K.},\ \bibinfo {year} {2012})\BibitemShut {NoStop}%
\bibitem [{\citenamefont {Yang}\ and\ \citenamefont {Nagaosa}(2014)}]{Yang14}%
  \BibitemOpen
  \bibfield  {author} {\bibinfo {author} {\bibfnamefont {B.}~\bibnamefont
  {Yang}}\ and\ \bibinfo {author} {\bibfnamefont {N.}~\bibnamefont {Nagaosa}},\
  }\bibfield  {booktitle} {\emph {\bibinfo {booktitle} {Classification of
  stable three-dimensional Dirac semimetals with nontrivial topology}},\ }\href
  {\doibase doi:10.1038/ncomms5898} {\bibfield  {journal} {\bibinfo  {journal}
  {Nat. Commun.}\ }\textbf {\bibinfo {volume} {5}},\ \bibinfo {pages} {4898}
  (\bibinfo {year} {2014})}\BibitemShut {NoStop}%
\bibitem [{\citenamefont {Burkov}(2016)}]{Burkov16}%
  \BibitemOpen
  \bibfield  {author} {\bibinfo {author} {\bibfnamefont {A.~A.}\ \bibnamefont
  {Burkov}},\ }\bibfield  {booktitle} {\emph {\bibinfo {booktitle} {Topological
  semimetals}},\ }\href {\doibase doi:10.1038/nmat4788} {\bibfield  {journal}
  {\bibinfo  {journal} {Nat. Mater.}\ }\textbf {\bibinfo {volume} {15}},\
  \bibinfo {pages} {1145} (\bibinfo {year} {2016})}\BibitemShut {NoStop}%
\bibitem [{\citenamefont {Armitage}\ \emph {et~al.}(2018)\citenamefont
  {Armitage}, \citenamefont {Mele},\ and\ \citenamefont
  {Vishwanath}}]{Armitage18}%
  \BibitemOpen
  \bibfield  {author} {\bibinfo {author} {\bibfnamefont {N.~P.}\ \bibnamefont
  {Armitage}}, \bibinfo {author} {\bibfnamefont {E.~J.}\ \bibnamefont {Mele}},
  \ and\ \bibinfo {author} {\bibfnamefont {A.}~\bibnamefont {Vishwanath}},\
  }\bibfield  {booktitle} {\emph {\bibinfo {booktitle} {Weyl and Dirac
  semimetals in three-dimensional solids}},\ }\href {\doibase
  10.1103/RevModPhys.90.015001} {\bibfield  {journal} {\bibinfo  {journal}
  {Rev. Mod. Phys.}\ }\textbf {\bibinfo {volume} {90}},\ \bibinfo {pages}
  {015001} (\bibinfo {year} {2018})}\BibitemShut {NoStop}%
\bibitem [{\citenamefont {Fang}\ \emph {et~al.}(2015)\citenamefont {Fang},
  \citenamefont {Chen}, \citenamefont {Kee},\ and\ \citenamefont
  {Fu}}]{Fang15}%
  \BibitemOpen
  \bibfield  {author} {\bibinfo {author} {\bibfnamefont {C.}~\bibnamefont
  {Fang}}, \bibinfo {author} {\bibfnamefont {Y.}~\bibnamefont {Chen}}, \bibinfo
  {author} {\bibfnamefont {H.-Y.}\ \bibnamefont {Kee}}, \ and\ \bibinfo
  {author} {\bibfnamefont {L.}~\bibnamefont {Fu}},\ }\bibfield  {booktitle}
  {\emph {\bibinfo {booktitle} {Topological nodal line semimetals with and
  without spin-orbital coupling}},\ }\href {\doibase
  10.1103/PhysRevB.92.081201} {\bibfield  {journal} {\bibinfo  {journal} {Phys.
  Rev. B}\ }\textbf {\bibinfo {volume} {92}},\ \bibinfo {pages} {081201}
  (\bibinfo {year} {2015})}\BibitemShut {NoStop}%
\bibitem [{\citenamefont {Liu}\ and\ \citenamefont {Balents}(2017)}]{Liu17}%
  \BibitemOpen
  \bibfield  {author} {\bibinfo {author} {\bibfnamefont {J.}~\bibnamefont
  {Liu}}\ and\ \bibinfo {author} {\bibfnamefont {L.}~\bibnamefont {Balents}},\
  }\bibfield  {booktitle} {\emph {\bibinfo {booktitle} {Correlation effects and
  quantum oscillations in topological nodal-loop semimetals}},\ }\href
  {\doibase 10.1103/PhysRevB.95.075426} {\bibfield  {journal} {\bibinfo
  {journal} {Phys. Rev. B}\ }\textbf {\bibinfo {volume} {95}},\ \bibinfo
  {pages} {075426} (\bibinfo {year} {2017})}\BibitemShut {NoStop}%
\bibitem [{\citenamefont {Schoop}\ \emph {et~al.}(2016)\citenamefont {Schoop},
  \citenamefont {Ali}, \citenamefont {Stra{\ss}er}, \citenamefont {Topp},
  \citenamefont {Varykhalov}, \citenamefont {Marchenko}, \citenamefont
  {Duppel}, \citenamefont {Parkin}, \citenamefont {Lotsch},\ and\ \citenamefont
  {Ast}}]{Schoop16}%
  \BibitemOpen
  \bibfield  {author} {\bibinfo {author} {\bibfnamefont {L.~M.}\ \bibnamefont
  {Schoop}}, \bibinfo {author} {\bibfnamefont {M.~N.}\ \bibnamefont {Ali}},
  \bibinfo {author} {\bibfnamefont {C.}~\bibnamefont {Stra{\ss}er}}, \bibinfo
  {author} {\bibfnamefont {A.}~\bibnamefont {Topp}}, \bibinfo {author}
  {\bibfnamefont {A.}~\bibnamefont {Varykhalov}}, \bibinfo {author}
  {\bibfnamefont {D.}~\bibnamefont {Marchenko}}, \bibinfo {author}
  {\bibfnamefont {V.}~\bibnamefont {Duppel}}, \bibinfo {author} {\bibfnamefont
  {S.~S.}\ \bibnamefont {Parkin}}, \bibinfo {author} {\bibfnamefont {B.~V.}\
  \bibnamefont {Lotsch}}, \ and\ \bibinfo {author} {\bibfnamefont {C.~R.}\
  \bibnamefont {Ast}},\ }\bibfield  {booktitle} {\emph {\bibinfo {booktitle}
  {Dirac cone protected by non-symmorphic symmetry and three-dimensional Dirac
  line node in ZrSiS}},\ }\href {\doibase 10.1038/ncomms11696} {\bibfield
  {journal} {\bibinfo  {journal} {Nat. Commun.}\ }\textbf {\bibinfo {volume}
  {7}},\ \bibinfo {pages} {11696} (\bibinfo {year} {2016})}\BibitemShut
  {NoStop}%
\bibitem [{\citenamefont {Neupane}\ \emph {et~al.}(2016)\citenamefont
  {Neupane}, \citenamefont {Belopolski}, \citenamefont {Hosen}, \citenamefont
  {Sanchez}, \citenamefont {Sankar}, \citenamefont {Szlawska}, \citenamefont
  {Xu}, \citenamefont {Dimitri}, \citenamefont {Dhakal}, \citenamefont
  {Maldonado}, \citenamefont {Oppeneer}, \citenamefont {Kaczorowski},
  \citenamefont {Chou}, \citenamefont {Hasan},\ and\ \citenamefont
  {Durakiewicz}}]{Neupane16}%
  \BibitemOpen
  \bibfield  {author} {\bibinfo {author} {\bibfnamefont {M.}~\bibnamefont
  {Neupane}}, \bibinfo {author} {\bibfnamefont {I.}~\bibnamefont {Belopolski}},
  \bibinfo {author} {\bibfnamefont {M.~M.}\ \bibnamefont {Hosen}}, \bibinfo
  {author} {\bibfnamefont {D.~S.}\ \bibnamefont {Sanchez}}, \bibinfo {author}
  {\bibfnamefont {R.}~\bibnamefont {Sankar}}, \bibinfo {author} {\bibfnamefont
  {M.}~\bibnamefont {Szlawska}}, \bibinfo {author} {\bibfnamefont {S.-Y.}\
  \bibnamefont {Xu}}, \bibinfo {author} {\bibfnamefont {K.}~\bibnamefont
  {Dimitri}}, \bibinfo {author} {\bibfnamefont {N.}~\bibnamefont {Dhakal}},
  \bibinfo {author} {\bibfnamefont {P.}~\bibnamefont {Maldonado}}, \bibinfo
  {author} {\bibfnamefont {P.~M.}\ \bibnamefont {Oppeneer}}, \bibinfo {author}
  {\bibfnamefont {D.}~\bibnamefont {Kaczorowski}}, \bibinfo {author}
  {\bibfnamefont {F.}~\bibnamefont {Chou}}, \bibinfo {author} {\bibfnamefont
  {M.~Z.}\ \bibnamefont {Hasan}}, \ and\ \bibinfo {author} {\bibfnamefont
  {T.}~\bibnamefont {Durakiewicz}},\ }\bibfield  {booktitle} {\emph {\bibinfo
  {booktitle} {Observation of topological nodal fermion semimetal phase in
  ZrSiS}},\ }\href {\doibase 10.1103/PhysRevB.93.201104} {\bibfield  {journal}
  {\bibinfo  {journal} {Phys. Rev. B}\ }\textbf {\bibinfo {volume} {93}},\
  \bibinfo {pages} {201104} (\bibinfo {year} {2016})}\BibitemShut {NoStop}%
\bibitem [{\citenamefont {Wang}\ \emph {et~al.}(2016)\citenamefont {Wang},
  \citenamefont {Pan}, \citenamefont {Gao}, \citenamefont {Yu}, \citenamefont
  {Jiang}, \citenamefont {Zhang}, \citenamefont {Zuo}, \citenamefont {Zhang},
  \citenamefont {Wei}, \citenamefont {Niu} \emph {et~al.}}]{Wang16}%
  \BibitemOpen
  \bibfield  {author} {\bibinfo {author} {\bibfnamefont {X.}~\bibnamefont
  {Wang}}, \bibinfo {author} {\bibfnamefont {X.}~\bibnamefont {Pan}}, \bibinfo
  {author} {\bibfnamefont {M.}~\bibnamefont {Gao}}, \bibinfo {author}
  {\bibfnamefont {J.}~\bibnamefont {Yu}}, \bibinfo {author} {\bibfnamefont
  {J.}~\bibnamefont {Jiang}}, \bibinfo {author} {\bibfnamefont
  {J.}~\bibnamefont {Zhang}}, \bibinfo {author} {\bibfnamefont
  {H.}~\bibnamefont {Zuo}}, \bibinfo {author} {\bibfnamefont {M.}~\bibnamefont
  {Zhang}}, \bibinfo {author} {\bibfnamefont {Z.}~\bibnamefont {Wei}}, \bibinfo
  {author} {\bibfnamefont {W.}~\bibnamefont {Niu}},  \emph {et~al.},\
  }\bibfield  {booktitle} {\emph {\bibinfo {booktitle} {Evidence of both
  surface and bulk Dirac bands and anisotropic nonsaturating magnetoresistance
  in ZrSiS}},\ }\href {\doibase 10.1002/aelm.201600228} {\bibfield  {journal}
  {\bibinfo  {journal} {Adv. Electron. Mater.}\ }\textbf {\bibinfo {volume}
  {2}},\ \bibinfo {pages} {1600228} (\bibinfo {year} {2016})}\BibitemShut
  {NoStop}%
\bibitem [{\citenamefont {Ali}\ \emph {et~al.}(2016)\citenamefont {Ali},
  \citenamefont {Schoop}, \citenamefont {Garg}, \citenamefont {Lippmann},
  \citenamefont {Lara}, \citenamefont {Lotsch},\ and\ \citenamefont
  {Parkin}}]{Ali16}%
  \BibitemOpen
  \bibfield  {author} {\bibinfo {author} {\bibfnamefont {M.~N.}\ \bibnamefont
  {Ali}}, \bibinfo {author} {\bibfnamefont {L.~M.}\ \bibnamefont {Schoop}},
  \bibinfo {author} {\bibfnamefont {C.}~\bibnamefont {Garg}}, \bibinfo {author}
  {\bibfnamefont {J.~M.}\ \bibnamefont {Lippmann}}, \bibinfo {author}
  {\bibfnamefont {E.}~\bibnamefont {Lara}}, \bibinfo {author} {\bibfnamefont
  {B.}~\bibnamefont {Lotsch}}, \ and\ \bibinfo {author} {\bibfnamefont {S.~S.}\
  \bibnamefont {Parkin}},\ }\bibfield  {booktitle} {\emph {\bibinfo {booktitle}
  {Butterfly magnetoresistance, quasi-2D Dirac Fermi surface and topological
  phase transition in ZrSiS}},\ }\href {\doibase 10.1126/sciadv.1601742}
  {\bibfield  {journal} {\bibinfo  {journal} {Sci. Adv.}\ }\textbf {\bibinfo
  {volume} {2}},\ \bibinfo {pages} {e1601742} (\bibinfo {year}
  {2016})}\BibitemShut {NoStop}%
\bibitem [{\citenamefont {Singha}\ \emph {et~al.}(2017)\citenamefont {Singha},
  \citenamefont {Pariari}, \citenamefont {Satpati},\ and\ \citenamefont
  {Mandal}}]{Singha17}%
  \BibitemOpen
  \bibfield  {author} {\bibinfo {author} {\bibfnamefont {R.}~\bibnamefont
  {Singha}}, \bibinfo {author} {\bibfnamefont {A.~K.}\ \bibnamefont {Pariari}},
  \bibinfo {author} {\bibfnamefont {B.}~\bibnamefont {Satpati}}, \ and\
  \bibinfo {author} {\bibfnamefont {P.}~\bibnamefont {Mandal}},\ }\bibfield
  {booktitle} {\emph {\bibinfo {booktitle} {Large nonsaturating
  magnetoresistance and signature of nondegenerate Dirac nodes in ZrSiS}},\
  }\href {\doibase 10.1073/pnas.1618004114} {\bibfield  {journal} {\bibinfo
  {journal} {Proc. Natl. Acad. Sci. U.S.A.}\ }\textbf {\bibinfo {volume}
  {114}},\ \bibinfo {pages} {2468} (\bibinfo {year} {2017})}\BibitemShut
  {NoStop}%
\bibitem [{\citenamefont {Matusiak}\ \emph {et~al.}(2017)\citenamefont
  {Matusiak}, \citenamefont {Cooper},\ and\ \citenamefont
  {Kaczorowski}}]{Matusiak17}%
  \BibitemOpen
  \bibfield  {author} {\bibinfo {author} {\bibfnamefont {M.}~\bibnamefont
  {Matusiak}}, \bibinfo {author} {\bibfnamefont {J.}~\bibnamefont {Cooper}}, \
  and\ \bibinfo {author} {\bibfnamefont {D.}~\bibnamefont {Kaczorowski}},\
  }\bibfield  {booktitle} {\emph {\bibinfo {booktitle} {Thermoelectric quantum
  oscillations in ZrSiS}},\ }\href {\doibase 10.1038/ncomms15219} {\bibfield
  {journal} {\bibinfo  {journal} {Nat. Commun.}\ }\textbf {\bibinfo {volume}
  {8}},\ \bibinfo {pages} {15219} (\bibinfo {year} {2017})}\BibitemShut
  {NoStop}%
\bibitem [{\citenamefont {Sankar}\ \emph {et~al.}(2017)\citenamefont {Sankar},
  \citenamefont {Peramaiyan}, \citenamefont {Muthuselvam}, \citenamefont
  {Butler}, \citenamefont {Dimitri}, \citenamefont {Neupane}, \citenamefont
  {Rao}, \citenamefont {Lin},\ and\ \citenamefont {Chou}}]{Sankar17}%
  \BibitemOpen
  \bibfield  {author} {\bibinfo {author} {\bibfnamefont {R.}~\bibnamefont
  {Sankar}}, \bibinfo {author} {\bibfnamefont {G.}~\bibnamefont {Peramaiyan}},
  \bibinfo {author} {\bibfnamefont {I.~P.}\ \bibnamefont {Muthuselvam}},
  \bibinfo {author} {\bibfnamefont {C.~J.}\ \bibnamefont {Butler}}, \bibinfo
  {author} {\bibfnamefont {K.}~\bibnamefont {Dimitri}}, \bibinfo {author}
  {\bibfnamefont {M.}~\bibnamefont {Neupane}}, \bibinfo {author} {\bibfnamefont
  {G.~N.}\ \bibnamefont {Rao}}, \bibinfo {author} {\bibfnamefont {M.-T.}\
  \bibnamefont {Lin}}, \ and\ \bibinfo {author} {\bibfnamefont
  {F.}~\bibnamefont {Chou}},\ }\bibfield  {booktitle} {\emph {\bibinfo
  {booktitle} {Crystal growth of Dirac semimetal ZrSiS with high
  magnetoresistance and mobility}},\ }\href {\doibase 10.1038/srep40603}
  {\bibfield  {journal} {\bibinfo  {journal} {Sci. Rep.}\ }\textbf {\bibinfo
  {volume} {7}},\ \bibinfo {pages} {40603} (\bibinfo {year}
  {2017})}\BibitemShut {NoStop}%
\bibitem [{\citenamefont {Lodge}\ \emph {et~al.}(2017)\citenamefont {Lodge},
  \citenamefont {Chang}, \citenamefont {Huang}, \citenamefont {Singh},
  \citenamefont {Hellerstedt}, \citenamefont {Edmonds}, \citenamefont
  {Kaczorowski}, \citenamefont {Hosen}, \citenamefont {Neupane}, \citenamefont
  {Lin} \emph {et~al.}}]{Lodge17}%
  \BibitemOpen
  \bibfield  {author} {\bibinfo {author} {\bibfnamefont {M.~S.}\ \bibnamefont
  {Lodge}}, \bibinfo {author} {\bibfnamefont {G.}~\bibnamefont {Chang}},
  \bibinfo {author} {\bibfnamefont {C.-Y.}\ \bibnamefont {Huang}}, \bibinfo
  {author} {\bibfnamefont {B.}~\bibnamefont {Singh}}, \bibinfo {author}
  {\bibfnamefont {J.}~\bibnamefont {Hellerstedt}}, \bibinfo {author}
  {\bibfnamefont {M.~T.}\ \bibnamefont {Edmonds}}, \bibinfo {author}
  {\bibfnamefont {D.}~\bibnamefont {Kaczorowski}}, \bibinfo {author}
  {\bibfnamefont {M.~M.}\ \bibnamefont {Hosen}}, \bibinfo {author}
  {\bibfnamefont {M.}~\bibnamefont {Neupane}}, \bibinfo {author} {\bibfnamefont
  {H.}~\bibnamefont {Lin}},  \emph {et~al.},\ }\bibfield  {booktitle} {\emph
  {\bibinfo {booktitle} {Observation of effective pseudospin scattering in
  ZrSiS}},\ }\href {\doibase 10.1021/acs.nanolett.7b02307} {\bibfield
  {journal} {\bibinfo  {journal} {Nano Lett.}\ }\textbf {\bibinfo {volume}
  {17}},\ \bibinfo {pages} {7213} (\bibinfo {year} {2017})}\BibitemShut
  {NoStop}%
\bibitem [{\citenamefont {Butler}\ \emph {et~al.}(2017)\citenamefont {Butler},
  \citenamefont {Wu}, \citenamefont {Hsing}, \citenamefont {Tseng},
  \citenamefont {Sankar}, \citenamefont {Wei}, \citenamefont {Chou},\ and\
  \citenamefont {Lin}}]{Butler17}%
  \BibitemOpen
  \bibfield  {author} {\bibinfo {author} {\bibfnamefont {C.~J.}\ \bibnamefont
  {Butler}}, \bibinfo {author} {\bibfnamefont {Y.-M.}\ \bibnamefont {Wu}},
  \bibinfo {author} {\bibfnamefont {C.-R.}\ \bibnamefont {Hsing}}, \bibinfo
  {author} {\bibfnamefont {Y.}~\bibnamefont {Tseng}}, \bibinfo {author}
  {\bibfnamefont {R.}~\bibnamefont {Sankar}}, \bibinfo {author} {\bibfnamefont
  {C.-M.}\ \bibnamefont {Wei}}, \bibinfo {author} {\bibfnamefont {F.-C.}\
  \bibnamefont {Chou}}, \ and\ \bibinfo {author} {\bibfnamefont {M.-T.}\
  \bibnamefont {Lin}},\ }\bibfield  {booktitle} {\emph {\bibinfo {booktitle}
  {Quasiparticle interference in ZrSiS: Strongly band-selective scattering
  depending on impurity lattice site}},\ }\href {\doibase
  10.1103/PhysRevB.96.195125} {\bibfield  {journal} {\bibinfo  {journal} {Phys.
  Rev. B}\ }\textbf {\bibinfo {volume} {96}},\ \bibinfo {pages} {195125}
  (\bibinfo {year} {2017})}\BibitemShut {NoStop}%
\bibitem [{\citenamefont {Topp}\ \emph {et~al.}(2017)\citenamefont {Topp},
  \citenamefont {Queiroz}, \citenamefont {Gr\"uneis}, \citenamefont
  {M\"uchler}, \citenamefont {Rost}, \citenamefont {Varykhalov}, \citenamefont
  {Marchenko}, \citenamefont {Krivenkov}, \citenamefont {Rodolakis},
  \citenamefont {McChesney}, \citenamefont {Lotsch}, \citenamefont {Schoop},\
  and\ \citenamefont {Ast}}]{Topp17}%
  \BibitemOpen
  \bibfield  {author} {\bibinfo {author} {\bibfnamefont {A.}~\bibnamefont
  {Topp}}, \bibinfo {author} {\bibfnamefont {R.}~\bibnamefont {Queiroz}},
  \bibinfo {author} {\bibfnamefont {A.}~\bibnamefont {Gr\"uneis}}, \bibinfo
  {author} {\bibfnamefont {L.}~\bibnamefont {M\"uchler}}, \bibinfo {author}
  {\bibfnamefont {A.~W.}\ \bibnamefont {Rost}}, \bibinfo {author}
  {\bibfnamefont {A.}~\bibnamefont {Varykhalov}}, \bibinfo {author}
  {\bibfnamefont {D.}~\bibnamefont {Marchenko}}, \bibinfo {author}
  {\bibfnamefont {M.}~\bibnamefont {Krivenkov}}, \bibinfo {author}
  {\bibfnamefont {F.}~\bibnamefont {Rodolakis}}, \bibinfo {author}
  {\bibfnamefont {J.~L.}\ \bibnamefont {McChesney}}, \bibinfo {author}
  {\bibfnamefont {B.~V.}\ \bibnamefont {Lotsch}}, \bibinfo {author}
  {\bibfnamefont {L.~M.}\ \bibnamefont {Schoop}}, \ and\ \bibinfo {author}
  {\bibfnamefont {C.~R.}\ \bibnamefont {Ast}},\ }\bibfield  {booktitle} {\emph
  {\bibinfo {booktitle} {Surface Floating 2D Bands in Layered Nonsymmorphic
  Semimetals: ZrSiS and Related Compounds}},\ }\href {\doibase
  10.1103/PhysRevX.7.041073} {\bibfield  {journal} {\bibinfo  {journal} {Phys.
  Rev. X}\ }\textbf {\bibinfo {volume} {7}},\ \bibinfo {pages} {041073}
  (\bibinfo {year} {2017})}\BibitemShut {NoStop}%
\bibitem [{\citenamefont {Hu}\ \emph {et~al.}(2017)\citenamefont {Hu},
  \citenamefont {Tang}, \citenamefont {Liu}, \citenamefont {Zhu}, \citenamefont
  {Wei},\ and\ \citenamefont {Mao}}]{Hu17}%
  \BibitemOpen
  \bibfield  {author} {\bibinfo {author} {\bibfnamefont {J.}~\bibnamefont
  {Hu}}, \bibinfo {author} {\bibfnamefont {Z.}~\bibnamefont {Tang}}, \bibinfo
  {author} {\bibfnamefont {J.}~\bibnamefont {Liu}}, \bibinfo {author}
  {\bibfnamefont {Y.}~\bibnamefont {Zhu}}, \bibinfo {author} {\bibfnamefont
  {J.}~\bibnamefont {Wei}}, \ and\ \bibinfo {author} {\bibfnamefont
  {Z.}~\bibnamefont {Mao}},\ }\bibfield  {booktitle} {\emph {\bibinfo
  {booktitle} {Nearly massless Dirac fermions and strong Zeeman splitting in
  the nodal-line semimetal ZrSiS probed by de Haas--van Alphen quantum
  oscillations}},\ }\href {\doibase 10.1103/PhysRevB.96.045127} {\bibfield
  {journal} {\bibinfo  {journal} {Phys. Rev. B}\ }\textbf {\bibinfo {volume}
  {96}},\ \bibinfo {pages} {045127} (\bibinfo {year} {2017})}\BibitemShut
  {NoStop}%
\bibitem [{\citenamefont {Schilling}\ \emph {et~al.}(2017)\citenamefont
  {Schilling}, \citenamefont {Schoop}, \citenamefont {Lotsch}, \citenamefont
  {Dressel},\ and\ \citenamefont {Pronin}}]{Schilling17}%
  \BibitemOpen
  \bibfield  {author} {\bibinfo {author} {\bibfnamefont {M.~B.}\ \bibnamefont
  {Schilling}}, \bibinfo {author} {\bibfnamefont {L.~M.}\ \bibnamefont
  {Schoop}}, \bibinfo {author} {\bibfnamefont {B.~V.}\ \bibnamefont {Lotsch}},
  \bibinfo {author} {\bibfnamefont {M.}~\bibnamefont {Dressel}}, \ and\
  \bibinfo {author} {\bibfnamefont {A.~V.}\ \bibnamefont {Pronin}},\ }\bibfield
   {booktitle} {\emph {\bibinfo {booktitle} {Flat Optical Conductivity in ZrSiS
  due to Two-Dimensional Dirac Bands}},\ }\href {\doibase
  10.1103/PhysRevLett.119.187401} {\bibfield  {journal} {\bibinfo  {journal}
  {Phys. Rev. Lett.}\ }\textbf {\bibinfo {volume} {119}},\ \bibinfo {pages}
  {187401} (\bibinfo {year} {2017})}\BibitemShut {NoStop}%
\bibitem [{\citenamefont {Pezzini}\ \emph {et~al.}(2018)\citenamefont
  {Pezzini}, \citenamefont {Van~Delft}, \citenamefont {Schoop}, \citenamefont
  {Lotsch}, \citenamefont {Carrington}, \citenamefont {Katsnelson},
  \citenamefont {Hussey},\ and\ \citenamefont {Wiedmann}}]{Pezzini18}%
  \BibitemOpen
  \bibfield  {author} {\bibinfo {author} {\bibfnamefont {S.}~\bibnamefont
  {Pezzini}}, \bibinfo {author} {\bibfnamefont {M.}~\bibnamefont {Van~Delft}},
  \bibinfo {author} {\bibfnamefont {L.}~\bibnamefont {Schoop}}, \bibinfo
  {author} {\bibfnamefont {B.}~\bibnamefont {Lotsch}}, \bibinfo {author}
  {\bibfnamefont {A.}~\bibnamefont {Carrington}}, \bibinfo {author}
  {\bibfnamefont {M.}~\bibnamefont {Katsnelson}}, \bibinfo {author}
  {\bibfnamefont {N.}~\bibnamefont {Hussey}}, \ and\ \bibinfo {author}
  {\bibfnamefont {S.}~\bibnamefont {Wiedmann}},\ }\bibfield  {booktitle} {\emph
  {\bibinfo {booktitle} {Unconventional mass enhancement around the Dirac nodal
  loop in ZrSiS}},\ }\href {\doibase 10.1038/nphys4306} {\bibfield  {journal}
  {\bibinfo  {journal} {Nat. Phys.}\ }\textbf {\bibinfo {volume} {14}},\
  \bibinfo {pages} {178} (\bibinfo {year} {2018})}\BibitemShut {NoStop}%
\bibitem [{\citenamefont {Rudenko}\ \emph {et~al.}(2018)\citenamefont
  {Rudenko}, \citenamefont {Stepanov}, \citenamefont {Lichtenstein},\ and\
  \citenamefont {Katsnelson}}]{Rudenko18}%
  \BibitemOpen
  \bibfield  {author} {\bibinfo {author} {\bibfnamefont {A.~N.}\ \bibnamefont
  {Rudenko}}, \bibinfo {author} {\bibfnamefont {E.~A.}\ \bibnamefont
  {Stepanov}}, \bibinfo {author} {\bibfnamefont {A.~I.}\ \bibnamefont
  {Lichtenstein}}, \ and\ \bibinfo {author} {\bibfnamefont {M.~I.}\
  \bibnamefont {Katsnelson}},\ }\bibfield  {booktitle} {\emph {\bibinfo
  {booktitle} {{Excitonic Instability and Pseudogap Formation in Nodal Line
  Semimetal ZrSiS}}},\ }\href {\doibase 10.1103/PhysRevLett.120.216401}
  {\bibfield  {journal} {\bibinfo  {journal} {Phys. Rev. Lett.}\ }\textbf
  {\bibinfo {volume} {120}},\ \bibinfo {pages} {216401} (\bibinfo {year}
  {2018})}\BibitemShut {NoStop}%
\bibitem [{\citenamefont {Scherer}\ \emph {et~al.}(2018)\citenamefont
  {Scherer}, \citenamefont {Honerkamp}, \citenamefont {Rudenko}, \citenamefont
  {Stepanov}, \citenamefont {Lichtenstein},\ and\ \citenamefont
  {Katsnelson}}]{Scherer18}%
  \BibitemOpen
  \bibfield  {author} {\bibinfo {author} {\bibfnamefont {M.~M.}\ \bibnamefont
  {Scherer}}, \bibinfo {author} {\bibfnamefont {C.}~\bibnamefont {Honerkamp}},
  \bibinfo {author} {\bibfnamefont {A.~N.}\ \bibnamefont {Rudenko}}, \bibinfo
  {author} {\bibfnamefont {E.~A.}\ \bibnamefont {Stepanov}}, \bibinfo {author}
  {\bibfnamefont {A.~I.}\ \bibnamefont {Lichtenstein}}, \ and\ \bibinfo
  {author} {\bibfnamefont {M.~I.}\ \bibnamefont {Katsnelson}},\ }\bibfield
  {booktitle} {\emph {\bibinfo {booktitle} {{Excitonic instability and
  unconventional pairing in the nodal-line materials ZrSiS and ZrSiSe}}},\
  }\href {\doibase 10.1103/PhysRevB.98.241112} {\bibfield  {journal} {\bibinfo
  {journal} {Phys. Rev. B}\ }\textbf {\bibinfo {volume} {98}},\ \bibinfo
  {pages} {241112} (\bibinfo {year} {2018})}\BibitemShut {NoStop}%
\bibitem [{\citenamefont {VanGennep}\ \emph {et~al.}(2019)\citenamefont
  {VanGennep}, \citenamefont {Paul}, \citenamefont {Yerger}, \citenamefont
  {Weir}, \citenamefont {Vohra},\ and\ \citenamefont {Hamlin}}]{VanGennep19}%
  \BibitemOpen
  \bibfield  {author} {\bibinfo {author} {\bibfnamefont {D.}~\bibnamefont
  {VanGennep}}, \bibinfo {author} {\bibfnamefont {T.~A.}\ \bibnamefont {Paul}},
  \bibinfo {author} {\bibfnamefont {C.~W.}\ \bibnamefont {Yerger}}, \bibinfo
  {author} {\bibfnamefont {S.~T.}\ \bibnamefont {Weir}}, \bibinfo {author}
  {\bibfnamefont {Y.~K.}\ \bibnamefont {Vohra}}, \ and\ \bibinfo {author}
  {\bibfnamefont {J.~J.}\ \bibnamefont {Hamlin}},\ }\bibfield  {booktitle}
  {\emph {\bibinfo {booktitle} {Possible pressure-induced topological quantum
  phase transition in the nodal line semimetal ZrSiS}},\ }\href {\doibase
  10.1103/PhysRevB.99.085204} {\bibfield  {journal} {\bibinfo  {journal} {Phys.
  Rev. B}\ }\textbf {\bibinfo {volume} {99}},\ \bibinfo {pages} {085204}
  (\bibinfo {year} {2019})}\BibitemShut {NoStop}%
\bibitem [{\citenamefont {Novak}\ \emph {et~al.}(2019)\citenamefont {Novak},
  \citenamefont {Zhang}, \citenamefont {Orbani\ifmmode~\acute{c}\else
  \'{c}\fi{}}, \citenamefont {Bili\ifmmode~\check{s}\else \v{s}\fi{}kov},
  \citenamefont {Eguchi}, \citenamefont {Paschen}, \citenamefont {Kimura},
  \citenamefont {Wang}, \citenamefont {Osada}, \citenamefont {Uchida},
  \citenamefont {Sato}, \citenamefont {Wu}, \citenamefont {Yazyev},\ and\
  \citenamefont {Kokanovi\ifmmode~\acute{c}\else \'{c}\fi{}}}]{Novak19}%
  \BibitemOpen
  \bibfield  {author} {\bibinfo {author} {\bibfnamefont {M.}~\bibnamefont
  {Novak}}, \bibinfo {author} {\bibfnamefont {S.~N.}\ \bibnamefont {Zhang}},
  \bibinfo {author} {\bibfnamefont {F.}~\bibnamefont
  {Orbani\ifmmode~\acute{c}\else \'{c}\fi{}}}, \bibinfo {author} {\bibfnamefont
  {N.}~\bibnamefont {Bili\ifmmode~\check{s}\else \v{s}\fi{}kov}}, \bibinfo
  {author} {\bibfnamefont {G.}~\bibnamefont {Eguchi}}, \bibinfo {author}
  {\bibfnamefont {S.}~\bibnamefont {Paschen}}, \bibinfo {author} {\bibfnamefont
  {A.}~\bibnamefont {Kimura}}, \bibinfo {author} {\bibfnamefont {X.~X.}\
  \bibnamefont {Wang}}, \bibinfo {author} {\bibfnamefont {T.}~\bibnamefont
  {Osada}}, \bibinfo {author} {\bibfnamefont {K.}~\bibnamefont {Uchida}},
  \bibinfo {author} {\bibfnamefont {M.}~\bibnamefont {Sato}}, \bibinfo {author}
  {\bibfnamefont {Q.~S.}\ \bibnamefont {Wu}}, \bibinfo {author} {\bibfnamefont
  {O.~V.}\ \bibnamefont {Yazyev}}, \ and\ \bibinfo {author} {\bibfnamefont
  {I.}~\bibnamefont {Kokanovi\ifmmode~\acute{c}\else \'{c}\fi{}}},\ }\bibfield
  {booktitle} {\emph {\bibinfo {booktitle} {Highly anisotropic interlayer
  magnetoresitance in ZrSiS nodal-line Dirac semimetal}},\ }\href {\doibase
  10.1103/PhysRevB.100.085137} {\bibfield  {journal} {\bibinfo  {journal}
  {Phys. Rev. B}\ }\textbf {\bibinfo {volume} {100}},\ \bibinfo {pages}
  {085137} (\bibinfo {year} {2019})}\BibitemShut {NoStop}%
\bibitem [{\citenamefont {Habe}\ and\ \citenamefont {Koshino}(2018)}]{Habe18}%
  \BibitemOpen
  \bibfield  {author} {\bibinfo {author} {\bibfnamefont {T.}~\bibnamefont
  {Habe}}\ and\ \bibinfo {author} {\bibfnamefont {M.}~\bibnamefont {Koshino}},\
  }\bibfield  {booktitle} {\emph {\bibinfo {booktitle} {{Dynamical conductivity
  in the topological nodal-line semimetal ZrSiS}}},\ }\href {\doibase
  10.1103/PhysRevB.98.125201} {\bibfield  {journal} {\bibinfo  {journal} {Phys.
  Rev. B}\ }\textbf {\bibinfo {volume} {98}},\ \bibinfo {pages} {125201}
  (\bibinfo {year} {2018})}\BibitemShut {NoStop}%
\bibitem [{\citenamefont {Zhou}\ \emph {et~al.}()\citenamefont {Zhou},
  \citenamefont {Rudenko},\ and\ \citenamefont {Yuan}}]{Zhou}%
  \BibitemOpen
  \bibfield  {author} {\bibinfo {author} {\bibfnamefont {W.}~\bibnamefont
  {Zhou}}, \bibinfo {author} {\bibfnamefont {A.~N.}\ \bibnamefont {Rudenko}}, \
  and\ \bibinfo {author} {\bibfnamefont {S.}~\bibnamefont {Yuan}},\ }\bibfield
  {booktitle} {\emph {\bibinfo {booktitle} {Effect of mechanical strain on the
  optical properties of nodal-line semimetal ZrSiS}},\ }\href@noop {}
  {\bibfield  {journal} {\bibinfo  {journal} {Adv. Electron. Mater.}\ }\textbf
  {\bibinfo {volume} {6}}}\BibitemShut {NoStop}%
\bibitem [{\citenamefont {Salmankurt}\ and\ \citenamefont
  {Duman}(2017)}]{Duman16}%
  \BibitemOpen
  \bibfield  {author} {\bibinfo {author} {\bibfnamefont {B.}~\bibnamefont
  {Salmankurt}}\ and\ \bibinfo {author} {\bibfnamefont {S.}~\bibnamefont
  {Duman}},\ }\bibfield  {booktitle} {\emph {\bibinfo {booktitle}
  {First-principles study of structural, mechanical, lattice dynamical and
  thermal properties of nodal-line semimetals ZrXY (X=Si,Ge; Y=S,Se)}},\ }\href
  {\doibase 10.1080/14786435.2016.1250967} {\bibfield  {journal} {\bibinfo
  {journal} {Philos. Mag.}\ }\textbf {\bibinfo {volume} {97}},\ \bibinfo
  {pages} {175} (\bibinfo {year} {2017})}\BibitemShut {NoStop}%
\bibitem [{\citenamefont {Zhou}\ \emph {et~al.}(2017)\citenamefont {Zhou},
  \citenamefont {Gao}, \citenamefont {Zhang}, \citenamefont {Fang},
  \citenamefont {Song}, \citenamefont {Hu}, \citenamefont {Stroppa},
  \citenamefont {Li}, \citenamefont {Wang}, \citenamefont {Ruan},\ and\
  \citenamefont {Ren}}]{Zhou17}%
  \BibitemOpen
  \bibfield  {author} {\bibinfo {author} {\bibfnamefont {W.}~\bibnamefont
  {Zhou}}, \bibinfo {author} {\bibfnamefont {H.}~\bibnamefont {Gao}}, \bibinfo
  {author} {\bibfnamefont {J.}~\bibnamefont {Zhang}}, \bibinfo {author}
  {\bibfnamefont {R.}~\bibnamefont {Fang}}, \bibinfo {author} {\bibfnamefont
  {H.}~\bibnamefont {Song}}, \bibinfo {author} {\bibfnamefont {T.}~\bibnamefont
  {Hu}}, \bibinfo {author} {\bibfnamefont {A.}~\bibnamefont {Stroppa}},
  \bibinfo {author} {\bibfnamefont {L.}~\bibnamefont {Li}}, \bibinfo {author}
  {\bibfnamefont {X.}~\bibnamefont {Wang}}, \bibinfo {author} {\bibfnamefont
  {S.}~\bibnamefont {Ruan}}, \ and\ \bibinfo {author} {\bibfnamefont
  {W.}~\bibnamefont {Ren}},\ }\bibfield  {booktitle} {\emph {\bibinfo
  {booktitle} {Lattice dynamics of Dirac node-line semimetal ZrSiS}},\ }\href
  {\doibase 10.1103/PhysRevB.96.064103} {\bibfield  {journal} {\bibinfo
  {journal} {Phys. Rev. B}\ }\textbf {\bibinfo {volume} {96}},\ \bibinfo
  {pages} {064103} (\bibinfo {year} {2017})}\BibitemShut {NoStop}%
\bibitem [{\citenamefont {Singha}\ \emph {et~al.}(2018)\citenamefont {Singha},
  \citenamefont {Samanta}, \citenamefont {Chatterjee}, \citenamefont {Pariari},
  \citenamefont {Majumdar}, \citenamefont {Satpati}, \citenamefont {Wang},
  \citenamefont {Singha},\ and\ \citenamefont {Mandal}}]{Singha18}%
  \BibitemOpen
  \bibfield  {author} {\bibinfo {author} {\bibfnamefont {R.}~\bibnamefont
  {Singha}}, \bibinfo {author} {\bibfnamefont {S.}~\bibnamefont {Samanta}},
  \bibinfo {author} {\bibfnamefont {S.}~\bibnamefont {Chatterjee}}, \bibinfo
  {author} {\bibfnamefont {A.}~\bibnamefont {Pariari}}, \bibinfo {author}
  {\bibfnamefont {D.}~\bibnamefont {Majumdar}}, \bibinfo {author}
  {\bibfnamefont {B.}~\bibnamefont {Satpati}}, \bibinfo {author} {\bibfnamefont
  {L.}~\bibnamefont {Wang}}, \bibinfo {author} {\bibfnamefont {A.}~\bibnamefont
  {Singha}}, \ and\ \bibinfo {author} {\bibfnamefont {P.}~\bibnamefont
  {Mandal}},\ }\bibfield  {booktitle} {\emph {\bibinfo {booktitle} {Probing
  lattice dynamics and electron-phonon coupling in the topological nodal-line
  semimetal ZrSiS}},\ }\href {\doibase 10.1103/PhysRevB.97.094112} {\bibfield
  {journal} {\bibinfo  {journal} {Phys. Rev. B}\ }\textbf {\bibinfo {volume}
  {97}},\ \bibinfo {pages} {094112} (\bibinfo {year} {2018})}\BibitemShut
  {NoStop}%
\bibitem [{\citenamefont {Xue}\ \emph {et~al.}(2019)\citenamefont {Xue},
  \citenamefont {Zhang}, \citenamefont {Yi}, \citenamefont {Zhang},
  \citenamefont {Jia}, \citenamefont {Santos}, \citenamefont {Fang},
  \citenamefont {Shi}, \citenamefont {Zhu},\ and\ \citenamefont {Guo}}]{Xue19}%
  \BibitemOpen
  \bibfield  {author} {\bibinfo {author} {\bibfnamefont {S.}~\bibnamefont
  {Xue}}, \bibinfo {author} {\bibfnamefont {T.}~\bibnamefont {Zhang}}, \bibinfo
  {author} {\bibfnamefont {C.}~\bibnamefont {Yi}}, \bibinfo {author}
  {\bibfnamefont {S.}~\bibnamefont {Zhang}}, \bibinfo {author} {\bibfnamefont
  {X.}~\bibnamefont {Jia}}, \bibinfo {author} {\bibfnamefont {L.~H.}\
  \bibnamefont {Santos}}, \bibinfo {author} {\bibfnamefont {C.}~\bibnamefont
  {Fang}}, \bibinfo {author} {\bibfnamefont {Y.}~\bibnamefont {Shi}}, \bibinfo
  {author} {\bibfnamefont {X.}~\bibnamefont {Zhu}}, \ and\ \bibinfo {author}
  {\bibfnamefont {J.}~\bibnamefont {Guo}},\ }\bibfield  {booktitle} {\emph
  {\bibinfo {booktitle} {Electron-phonon Coupling and Kohn Anomaly due to the
  Floating 2D Electronic Bands on the Surface of ZrSiS}},\ }\href {\doibase
  10.1103/PhysRevB.100.195409} {\bibfield  {journal} {\bibinfo  {journal}
  {Phys. Rev. B}\ }\textbf {\bibinfo {volume} {100}},\ \bibinfo {pages}
  {195409} (\bibinfo {year} {2019})}\BibitemShut {NoStop}%
\bibitem [{\citenamefont {Syzranov}\ and\ \citenamefont
  {Skinner}(2017)}]{Syzranov17}%
  \BibitemOpen
  \bibfield  {author} {\bibinfo {author} {\bibfnamefont {S.~V.}\ \bibnamefont
  {Syzranov}}\ and\ \bibinfo {author} {\bibfnamefont {B.}~\bibnamefont
  {Skinner}},\ }\bibfield  {booktitle} {\emph {\bibinfo {booktitle} {Electron
  transport in nodal-line semimetals}},\ }\href {\doibase
  10.1103/PhysRevB.96.161105} {\bibfield  {journal} {\bibinfo  {journal} {Phys.
  Rev. B}\ }\textbf {\bibinfo {volume} {96}},\ \bibinfo {pages} {161105}
  (\bibinfo {year} {2017})}\BibitemShut {NoStop}%
\bibitem [{\citenamefont {Shirer}\ \emph {et~al.}(2019)\citenamefont {Shirer},
  \citenamefont {Modic}, \citenamefont {Zimmerling}, \citenamefont {Bachmann},
  \citenamefont {König}, \citenamefont {Moll}, \citenamefont {Schoop},\ and\
  \citenamefont {Mackenzie}}]{Shirer19}%
  \BibitemOpen
  \bibfield  {author} {\bibinfo {author} {\bibfnamefont {K.~R.}\ \bibnamefont
  {Shirer}}, \bibinfo {author} {\bibfnamefont {K.~A.}\ \bibnamefont {Modic}},
  \bibinfo {author} {\bibfnamefont {T.}~\bibnamefont {Zimmerling}}, \bibinfo
  {author} {\bibfnamefont {M.~D.}\ \bibnamefont {Bachmann}}, \bibinfo {author}
  {\bibfnamefont {M.}~\bibnamefont {König}}, \bibinfo {author} {\bibfnamefont
  {P.~J.~W.}\ \bibnamefont {Moll}}, \bibinfo {author} {\bibfnamefont
  {L.}~\bibnamefont {Schoop}}, \ and\ \bibinfo {author} {\bibfnamefont {A.~P.}\
  \bibnamefont {Mackenzie}},\ }\bibfield  {booktitle} {\emph {\bibinfo
  {booktitle} {Out-of-plane transport in ZrSiS and ZrSiSe microstructures}},\
  }\href {\doibase 10.1063/1.5124568} {\bibfield  {journal} {\bibinfo
  {journal} {APL Mater.}\ }\textbf {\bibinfo {volume} {7}},\ \bibinfo {pages}
  {101116} (\bibinfo {year} {2019})}\BibitemShut {NoStop}%
\bibitem [{\citenamefont {Ziman}(1960)}]{Ziman}%
  \BibitemOpen
  \bibfield  {author} {\bibinfo {author} {\bibfnamefont {Z.~M.}\ \bibnamefont
  {Ziman}},\ }\href@noop {} {\emph {\bibinfo {title} {Electrons and Phonons:
  The Theory of Transport Phenomena in Solids}}}\ (\bibinfo  {publisher}
  {Oxford University Press, Oxford},\ \bibinfo {year} {1960})\BibitemShut
  {NoStop}%
\bibitem [{\citenamefont {Grimvall}(1999)}]{Grimvall}%
  \BibitemOpen
  \bibfield  {author} {\bibinfo {author} {\bibfnamefont {G.}~\bibnamefont
  {Grimvall}},\ }\href@noop {} {\emph {\bibinfo {title} {Thermophysical
  Properties of Materials}}}\ (\bibinfo  {publisher} {Elsevier Science,
  Amsterdam},\ \bibinfo {year} {1999})\BibitemShut {NoStop}%
\bibitem [{\citenamefont {Ponc\'e}\ \emph {et~al.}(2018)\citenamefont
  {Ponc\'e}, \citenamefont {Margine},\ and\ \citenamefont
  {Giustino}}]{Ponce2018}%
  \BibitemOpen
  \bibfield  {author} {\bibinfo {author} {\bibfnamefont {S.}~\bibnamefont
  {Ponc\'e}}, \bibinfo {author} {\bibfnamefont {E.~R.}\ \bibnamefont
  {Margine}}, \ and\ \bibinfo {author} {\bibfnamefont {F.}~\bibnamefont
  {Giustino}},\ }\bibfield  {booktitle} {\emph {\bibinfo {booktitle} {Towards
  predictive many-body calculations of phonon-limited carrier mobilities in
  semiconductors}},\ }\href {\doibase 10.1103/PhysRevB.97.121201} {\bibfield
  {journal} {\bibinfo  {journal} {Phys. Rev. B}\ }\textbf {\bibinfo {volume}
  {97}},\ \bibinfo {pages} {121201} (\bibinfo {year} {2018})}\BibitemShut
  {NoStop}%
\bibitem [{\citenamefont {Migdal}()}]{Migdal1958}%
  \BibitemOpen
  \bibfield  {author} {\bibinfo {author} {\bibfnamefont {A.~B.}\ \bibnamefont
  {Migdal}},\ }\href@noop {} {\bibinfo  {journal} {Sov. Phys. JETP}\ ,\
  \bibinfo {pages} {996}}\BibitemShut {NoStop}%
\bibitem [{\citenamefont {Marzari}\ \emph {et~al.}(2012)\citenamefont
  {Marzari}, \citenamefont {Mostofi}, \citenamefont {Yates}, \citenamefont
  {Souza},\ and\ \citenamefont {Vanderbilt}}]{Marzari2012}%
  \BibitemOpen
\bibfield  {journal} {  }\bibfield  {author} {\bibinfo {author} {\bibfnamefont
  {N.}~\bibnamefont {Marzari}}, \bibinfo {author} {\bibfnamefont {A.~A.}\
  \bibnamefont {Mostofi}}, \bibinfo {author} {\bibfnamefont {J.~R.}\
  \bibnamefont {Yates}}, \bibinfo {author} {\bibfnamefont {I.}~\bibnamefont
  {Souza}}, \ and\ \bibinfo {author} {\bibfnamefont {D.}~\bibnamefont
  {Vanderbilt}},\ }\bibfield  {booktitle} {\emph {\bibinfo {booktitle}
  {{Maximally localized Wannier functions: Theory and applications}}},\ }\href
  {\doibase 10.1103/RevModPhys.84.1419} {\bibfield  {journal} {\bibinfo
  {journal} {Rev. Mod. Phys.}\ }\textbf {\bibinfo {volume} {84}},\ \bibinfo
  {pages} {1419} (\bibinfo {year} {2012})}\BibitemShut {NoStop}%
\bibitem [{\citenamefont {Giannozzi}\ \emph {et~al.}(2017)\citenamefont
  {Giannozzi}, \citenamefont {Andreussi}, \citenamefont {Brumme}, \citenamefont
  {Bunau}, \citenamefont {{Buongiorno Nardelli}}, \citenamefont {Calandra},
  \citenamefont {Car}, \citenamefont {Cavazzoni}, \citenamefont {Ceresoli},
  \citenamefont {Cococcioni}, \citenamefont {Colonna}, \citenamefont
  {Carnimeo}, \citenamefont {{Dal Corso}}, \citenamefont {de~Gironcoli},
  \citenamefont {Delugas}, \citenamefont {DiStasio}, \citenamefont {Ferretti},
  \citenamefont {Floris}, \citenamefont {Fratesi}, \citenamefont {Fugallo},
  \citenamefont {Gebauer}, \citenamefont {Gerstmann}, \citenamefont {Giustino},
  \citenamefont {Gorni}, \citenamefont {Jia}, \citenamefont {Kawamura},
  \citenamefont {Ko}, \citenamefont {Kokalj}, \citenamefont
  {K{\"{u}}{\c{c}}{\"{u}}kbenli}, \citenamefont {Lazzeri}, \citenamefont
  {Marsili}, \citenamefont {Marzari}, \citenamefont {Mauri}, \citenamefont
  {Nguyen}, \citenamefont {Nguyen}, \citenamefont {Otero-de-la Roza},
  \citenamefont {Paulatto}, \citenamefont {Ponc{\'{e}}}, \citenamefont {Rocca},
  \citenamefont {Sabatini}, \citenamefont {Santra}, \citenamefont {Schlipf},
  \citenamefont {Seitsonen}, \citenamefont {Smogunov}, \citenamefont {Timrov},
  \citenamefont {Thonhauser}, \citenamefont {Umari}, \citenamefont {Vast},
  \citenamefont {Wu},\ and\ \citenamefont {Baroni}}]{Giannozzi2017}%
  \BibitemOpen
  \bibfield  {author} {\bibinfo {author} {\bibfnamefont {P.}~\bibnamefont
  {Giannozzi}}, \bibinfo {author} {\bibfnamefont {O.}~\bibnamefont
  {Andreussi}}, \bibinfo {author} {\bibfnamefont {T.}~\bibnamefont {Brumme}},
  \bibinfo {author} {\bibfnamefont {O.}~\bibnamefont {Bunau}}, \bibinfo
  {author} {\bibfnamefont {M.}~\bibnamefont {{Buongiorno Nardelli}}}, \bibinfo
  {author} {\bibfnamefont {M.}~\bibnamefont {Calandra}}, \bibinfo {author}
  {\bibfnamefont {R.}~\bibnamefont {Car}}, \bibinfo {author} {\bibfnamefont
  {C.}~\bibnamefont {Cavazzoni}}, \bibinfo {author} {\bibfnamefont
  {D.}~\bibnamefont {Ceresoli}}, \bibinfo {author} {\bibfnamefont
  {M.}~\bibnamefont {Cococcioni}}, \bibinfo {author} {\bibfnamefont
  {N.}~\bibnamefont {Colonna}}, \bibinfo {author} {\bibfnamefont
  {I.}~\bibnamefont {Carnimeo}}, \bibinfo {author} {\bibfnamefont
  {A.}~\bibnamefont {{Dal Corso}}}, \bibinfo {author} {\bibfnamefont
  {S.}~\bibnamefont {de~Gironcoli}}, \bibinfo {author} {\bibfnamefont
  {P.}~\bibnamefont {Delugas}}, \bibinfo {author} {\bibfnamefont {R.~A.}\
  \bibnamefont {DiStasio}}, \bibinfo {author} {\bibfnamefont {A.}~\bibnamefont
  {Ferretti}}, \bibinfo {author} {\bibfnamefont {A.}~\bibnamefont {Floris}},
  \bibinfo {author} {\bibfnamefont {G.}~\bibnamefont {Fratesi}}, \bibinfo
  {author} {\bibfnamefont {G.}~\bibnamefont {Fugallo}}, \bibinfo {author}
  {\bibfnamefont {R.}~\bibnamefont {Gebauer}}, \bibinfo {author} {\bibfnamefont
  {U.}~\bibnamefont {Gerstmann}}, \bibinfo {author} {\bibfnamefont
  {F.}~\bibnamefont {Giustino}}, \bibinfo {author} {\bibfnamefont
  {T.}~\bibnamefont {Gorni}}, \bibinfo {author} {\bibfnamefont
  {J.}~\bibnamefont {Jia}}, \bibinfo {author} {\bibfnamefont {M.}~\bibnamefont
  {Kawamura}}, \bibinfo {author} {\bibfnamefont {H.-Y.}\ \bibnamefont {Ko}},
  \bibinfo {author} {\bibfnamefont {A.}~\bibnamefont {Kokalj}}, \bibinfo
  {author} {\bibfnamefont {E.}~\bibnamefont {K{\"{u}}{\c{c}}{\"{u}}kbenli}},
  \bibinfo {author} {\bibfnamefont {M.}~\bibnamefont {Lazzeri}}, \bibinfo
  {author} {\bibfnamefont {M.}~\bibnamefont {Marsili}}, \bibinfo {author}
  {\bibfnamefont {N.}~\bibnamefont {Marzari}}, \bibinfo {author} {\bibfnamefont
  {F.}~\bibnamefont {Mauri}}, \bibinfo {author} {\bibfnamefont {N.~L.}\
  \bibnamefont {Nguyen}}, \bibinfo {author} {\bibfnamefont {H.-V.}\
  \bibnamefont {Nguyen}}, \bibinfo {author} {\bibfnamefont {A.}~\bibnamefont
  {Otero-de-la Roza}}, \bibinfo {author} {\bibfnamefont {L.}~\bibnamefont
  {Paulatto}}, \bibinfo {author} {\bibfnamefont {S.}~\bibnamefont
  {Ponc{\'{e}}}}, \bibinfo {author} {\bibfnamefont {D.}~\bibnamefont {Rocca}},
  \bibinfo {author} {\bibfnamefont {R.}~\bibnamefont {Sabatini}}, \bibinfo
  {author} {\bibfnamefont {B.}~\bibnamefont {Santra}}, \bibinfo {author}
  {\bibfnamefont {M.}~\bibnamefont {Schlipf}}, \bibinfo {author} {\bibfnamefont
  {A.~P.}\ \bibnamefont {Seitsonen}}, \bibinfo {author} {\bibfnamefont
  {A.}~\bibnamefont {Smogunov}}, \bibinfo {author} {\bibfnamefont
  {I.}~\bibnamefont {Timrov}}, \bibinfo {author} {\bibfnamefont
  {T.}~\bibnamefont {Thonhauser}}, \bibinfo {author} {\bibfnamefont
  {P.}~\bibnamefont {Umari}}, \bibinfo {author} {\bibfnamefont
  {N.}~\bibnamefont {Vast}}, \bibinfo {author} {\bibfnamefont {X.}~\bibnamefont
  {Wu}}, \ and\ \bibinfo {author} {\bibfnamefont {S.}~\bibnamefont {Baroni}},\
  }\bibfield  {booktitle} {\emph {\bibinfo {booktitle} {{Advanced capabilities
  for materials modeling with Quantum ESPRESSO}}},\ }\href {\doibase
  10.1088/1361-648X/aa8f79} {\bibfield  {journal} {\bibinfo  {journal} {J.
  Phys. Condens. Matter}\ }\textbf {\bibinfo {volume} {29}},\ \bibinfo {pages}
  {465901} (\bibinfo {year} {2017})}\BibitemShut {NoStop}%
\bibitem [{\citenamefont {Goedecker}\ \emph {et~al.}(1996)\citenamefont
  {Goedecker}, \citenamefont {Teter},\ and\ \citenamefont {Hutter}}]{gth}%
  \BibitemOpen
  \bibfield  {author} {\bibinfo {author} {\bibfnamefont {S.}~\bibnamefont
  {Goedecker}}, \bibinfo {author} {\bibfnamefont {M.}~\bibnamefont {Teter}}, \
  and\ \bibinfo {author} {\bibfnamefont {J.}~\bibnamefont {Hutter}},\
  }\bibfield  {booktitle} {\emph {\bibinfo {booktitle} {Separable dual-space
  Gaussian pseudopotentials}},\ }\href {\doibase 10.1103/PhysRevB.54.1703}
  {\bibfield  {journal} {\bibinfo  {journal} {Phys. Rev. B}\ }\textbf {\bibinfo
  {volume} {54}},\ \bibinfo {pages} {1703} (\bibinfo {year}
  {1996})}\BibitemShut {NoStop}%
\bibitem [{\citenamefont {Hartwigsen}\ \emph {et~al.}(1998)\citenamefont
  {Hartwigsen}, \citenamefont {Goedecker},\ and\ \citenamefont {Hutter}}]{hgh}%
  \BibitemOpen
  \bibfield  {author} {\bibinfo {author} {\bibfnamefont {C.}~\bibnamefont
  {Hartwigsen}}, \bibinfo {author} {\bibfnamefont {S.}~\bibnamefont
  {Goedecker}}, \ and\ \bibinfo {author} {\bibfnamefont {J.}~\bibnamefont
  {Hutter}},\ }\bibfield  {booktitle} {\emph {\bibinfo {booktitle}
  {Relativistic separable dual-space Gaussian pseudopotentials from H to Rn}},\
  }\href {\doibase 10.1103/PhysRevB.58.3641} {\bibfield  {journal} {\bibinfo
  {journal} {Phys. Rev. B}\ }\textbf {\bibinfo {volume} {58}},\ \bibinfo
  {pages} {3641} (\bibinfo {year} {1998})}\BibitemShut {NoStop}%
\bibitem [{\citenamefont {Perdew}\ and\ \citenamefont {Zunger}(1981)}]{lda-pz}%
  \BibitemOpen
  \bibfield  {author} {\bibinfo {author} {\bibfnamefont {J.~P.}\ \bibnamefont
  {Perdew}}\ and\ \bibinfo {author} {\bibfnamefont {A.}~\bibnamefont
  {Zunger}},\ }\bibfield  {booktitle} {\emph {\bibinfo {booktitle}
  {Self-interaction correction to density-functional approximations for
  many-electron systems}},\ }\href {\doibase 10.1103/PhysRevB.23.5048}
  {\bibfield  {journal} {\bibinfo  {journal} {Phys. Rev. B}\ }\textbf {\bibinfo
  {volume} {23}},\ \bibinfo {pages} {5048} (\bibinfo {year}
  {1981})}\BibitemShut {NoStop}%
\bibitem [{\citenamefont {Monkhorst}\ and\ \citenamefont
  {Pack}(1976)}]{Monkhorst1976}%
  \BibitemOpen
  \bibfield  {author} {\bibinfo {author} {\bibfnamefont {H.~J.}\ \bibnamefont
  {Monkhorst}}\ and\ \bibinfo {author} {\bibfnamefont {J.~D.}\ \bibnamefont
  {Pack}},\ }\bibfield  {booktitle} {\emph {\bibinfo {booktitle} {{Special
  points for Brillouin-zone integrations}}},\ }\href {\doibase
  10.1103/PhysRevB.13.5188} {\bibfield  {journal} {\bibinfo  {journal} {Phys.
  Rev. B}\ }\textbf {\bibinfo {volume} {13}},\ \bibinfo {pages} {5188}
  (\bibinfo {year} {1976})}\BibitemShut {NoStop}%
\bibitem [{\citenamefont {Baroni}\ \emph {et~al.}(2001)\citenamefont {Baroni},
  \citenamefont {de~Gironcoli}, \citenamefont {Dal~Corso},\ and\ \citenamefont
  {Giannozzi}}]{BaroniRMP}%
  \BibitemOpen
  \bibfield  {author} {\bibinfo {author} {\bibfnamefont {S.}~\bibnamefont
  {Baroni}}, \bibinfo {author} {\bibfnamefont {S.}~\bibnamefont
  {de~Gironcoli}}, \bibinfo {author} {\bibfnamefont {A.}~\bibnamefont
  {Dal~Corso}}, \ and\ \bibinfo {author} {\bibfnamefont {P.}~\bibnamefont
  {Giannozzi}},\ }\bibfield  {booktitle} {\emph {\bibinfo {booktitle} {Phonons
  and related crystal properties from density-functional perturbation
  theory}},\ }\href {\doibase 10.1103/RevModPhys.73.515} {\bibfield  {journal}
  {\bibinfo  {journal} {Rev. Mod. Phys.}\ }\textbf {\bibinfo {volume} {73}},\
  \bibinfo {pages} {515} (\bibinfo {year} {2001})}\BibitemShut {NoStop}%
\bibitem [{SI()}]{SI}%
  \BibitemOpen
  \href@noop {} {\ }\bibinfo {note} {See Supplemental Material for calculations
  demonstrating insignificant role of spin-orbit coupling in ZrSiS for the
  purposes of this paper, and for convergence tests demonstrating sufficient
  numerical accuracy of the Brillouin-zone integrals.}\BibitemShut {Stop}%
\bibitem [{\citenamefont {Giustino}(2017)}]{Giustino2017}%
  \BibitemOpen
  \bibfield  {author} {\bibinfo {author} {\bibfnamefont {F.}~\bibnamefont
  {Giustino}},\ }\bibfield  {booktitle} {\emph {\bibinfo {booktitle}
  {{Electron-phonon interactions from first principles}}},\ }\href {\doibase
  10.1103/RevModPhys.89.015003} {\bibfield  {journal} {\bibinfo  {journal}
  {Rev. Mod. Phys.}\ }\textbf {\bibinfo {volume} {89}},\ \bibinfo {pages}
  {015003} (\bibinfo {year} {2017})}\BibitemShut {NoStop}%
\bibitem [{\citenamefont {Mostofi}\ \emph {et~al.}(2014)\citenamefont
  {Mostofi}, \citenamefont {Yates}, \citenamefont {Pizzi}, \citenamefont {Lee},
  \citenamefont {Souza}, \citenamefont {Vanderbilt},\ and\ \citenamefont
  {Marzari}}]{wannier90}%
  \BibitemOpen
  \bibfield  {author} {\bibinfo {author} {\bibfnamefont {A.~A.}\ \bibnamefont
  {Mostofi}}, \bibinfo {author} {\bibfnamefont {J.~R.}\ \bibnamefont {Yates}},
  \bibinfo {author} {\bibfnamefont {G.}~\bibnamefont {Pizzi}}, \bibinfo
  {author} {\bibfnamefont {Y.-S.}\ \bibnamefont {Lee}}, \bibinfo {author}
  {\bibfnamefont {I.}~\bibnamefont {Souza}}, \bibinfo {author} {\bibfnamefont
  {D.}~\bibnamefont {Vanderbilt}}, \ and\ \bibinfo {author} {\bibfnamefont
  {N.}~\bibnamefont {Marzari}},\ }\bibfield  {booktitle} {\emph {\bibinfo
  {booktitle} {{An updated version of wannier90: A tool for obtaining
  maximally-localised Wannier functions}}},\ }\href {\doibase
  10.1016/j.cpc.2014.05.003} {\bibfield  {journal} {\bibinfo  {journal} {Comp.
  Phys. Commun.}\ }\textbf {\bibinfo {volume} {185}},\ \bibinfo {pages} {2309}
  (\bibinfo {year} {2014})}\BibitemShut {NoStop}%
\bibitem [{\citenamefont {Ponc{\'{e}}}\ \emph {et~al.}(2016)\citenamefont
  {Ponc{\'{e}}}, \citenamefont {Margine}, \citenamefont {Verdi},\ and\
  \citenamefont {Giustino}}]{Ponce2016}%
  \BibitemOpen
  \bibfield  {author} {\bibinfo {author} {\bibfnamefont {S.}~\bibnamefont
  {Ponc{\'{e}}}}, \bibinfo {author} {\bibfnamefont {E.~R.}\ \bibnamefont
  {Margine}}, \bibinfo {author} {\bibfnamefont {C.}~\bibnamefont {Verdi}}, \
  and\ \bibinfo {author} {\bibfnamefont {F.}~\bibnamefont {Giustino}},\
  }\bibfield  {booktitle} {\emph {\bibinfo {booktitle} {{EPW: Electron-phonon
  coupling, transport and superconducting properties using maximally localized
  Wannier functions}}},\ }\href {\doibase 10.1016/J.CPC.2016.07.028} {\bibfield
   {journal} {\bibinfo  {journal} {Comput. Phys. Commun.}\ }\textbf {\bibinfo
  {volume} {209}},\ \bibinfo {pages} {116} (\bibinfo {year}
  {2016})}\BibitemShut {NoStop}%
\bibitem [{\citenamefont {Kokalj}(1999)}]{xcrysden}%
  \BibitemOpen
  \bibfield  {author} {\bibinfo {author} {\bibfnamefont {A.}~\bibnamefont
  {Kokalj}},\ }\bibfield  {booktitle} {\emph {\bibinfo {booktitle} {XCrySDen -
  a new program for displaying crystalline structures and electron
  densities}},\ }\href {\doibase https://doi.org/10.1016/S1093-3263(99)00028-5}
  {\bibfield  {journal} {\bibinfo  {journal} {J. Mol. Graph. Model.}\ }\textbf
  {\bibinfo {volume} {17}},\ \bibinfo {pages} {176 } (\bibinfo {year}
  {1999})}\BibitemShut {NoStop}%
\bibitem [{\citenamefont {Allen}\ and\ \citenamefont
  {Dynes}(1975)}]{AllenDynes}%
  \BibitemOpen
  \bibfield  {author} {\bibinfo {author} {\bibfnamefont {P.~B.}\ \bibnamefont
  {Allen}}\ and\ \bibinfo {author} {\bibfnamefont {R.~C.}\ \bibnamefont
  {Dynes}},\ }\bibfield  {booktitle} {\emph {\bibinfo {booktitle} {Transition
  temperature of strong-coupled superconductors reanalyzed}},\ }\href {\doibase
  10.1103/PhysRevB.12.905} {\bibfield  {journal} {\bibinfo  {journal} {Phys.
  Rev. B}\ }\textbf {\bibinfo {volume} {12}},\ \bibinfo {pages} {905} (\bibinfo
  {year} {1975})}\BibitemShut {NoStop}%
\bibitem [{\citenamefont {Mustafa}\ \emph {et~al.}(2016)\citenamefont
  {Mustafa}, \citenamefont {Bernardi}, \citenamefont {Neaton},\ and\
  \citenamefont {Louie}}]{Louie16}%
  \BibitemOpen
  \bibfield  {author} {\bibinfo {author} {\bibfnamefont {J.~I.}\ \bibnamefont
  {Mustafa}}, \bibinfo {author} {\bibfnamefont {M.}~\bibnamefont {Bernardi}},
  \bibinfo {author} {\bibfnamefont {J.~B.}\ \bibnamefont {Neaton}}, \ and\
  \bibinfo {author} {\bibfnamefont {S.~G.}\ \bibnamefont {Louie}},\ }\bibfield
  {booktitle} {\emph {\bibinfo {booktitle} {Ab initio electronic relaxation
  times and transport in noble metals}},\ }\href {\doibase
  10.1103/PhysRevB.94.155105} {\bibfield  {journal} {\bibinfo  {journal} {Phys.
  Rev. B}\ }\textbf {\bibinfo {volume} {94}},\ \bibinfo {pages} {155105}
  (\bibinfo {year} {2016})}\BibitemShut {NoStop}%
\bibitem [{\citenamefont {Boukhvalov}\ \emph {et~al.}(2019)\citenamefont
  {Boukhvalov}, \citenamefont {Edla}, \citenamefont {Cupolillo}, \citenamefont
  {Fabio}, \citenamefont {Sankar}, \citenamefont {Zhu}, \citenamefont {Mao},
  \citenamefont {Hu}, \citenamefont {Torelli}, \citenamefont {Chiarello},
  \citenamefont {Ottaviano},\ and\ \citenamefont {Politano}}]{Boukhvalov19}%
  \BibitemOpen
  \bibfield  {author} {\bibinfo {author} {\bibfnamefont {D.~W.}\ \bibnamefont
  {Boukhvalov}}, \bibinfo {author} {\bibfnamefont {R.}~\bibnamefont {Edla}},
  \bibinfo {author} {\bibfnamefont {A.}~\bibnamefont {Cupolillo}}, \bibinfo
  {author} {\bibfnamefont {V.}~\bibnamefont {Fabio}}, \bibinfo {author}
  {\bibfnamefont {R.}~\bibnamefont {Sankar}}, \bibinfo {author} {\bibfnamefont
  {Y.}~\bibnamefont {Zhu}}, \bibinfo {author} {\bibfnamefont {Z.}~\bibnamefont
  {Mao}}, \bibinfo {author} {\bibfnamefont {J.}~\bibnamefont {Hu}}, \bibinfo
  {author} {\bibfnamefont {P.}~\bibnamefont {Torelli}}, \bibinfo {author}
  {\bibfnamefont {G.}~\bibnamefont {Chiarello}}, \bibinfo {author}
  {\bibfnamefont {L.}~\bibnamefont {Ottaviano}}, \ and\ \bibinfo {author}
  {\bibfnamefont {A.}~\bibnamefont {Politano}},\ }\bibfield  {booktitle} {\emph
  {\bibinfo {booktitle} {Surface Instability and Chemical Reactivity of ZrSiS
  and ZrSiSe Nodal-Line Semimetals}},\ }\href {\doibase 10.1002/adfm.201900438}
  {\bibfield  {journal} {\bibinfo  {journal} {Adv. Funct. Mater.}\ }\textbf
  {\bibinfo {volume} {29}},\ \bibinfo {pages} {1900438} (\bibinfo {year}
  {2019})}\BibitemShut {NoStop}%
\bibitem [{\citenamefont {Trevisanutto}\ \emph {et~al.}(2008)\citenamefont
  {Trevisanutto}, \citenamefont {Giorgetti}, \citenamefont {Reining},
  \citenamefont {Ladisa},\ and\ \citenamefont {Olevano}}]{Trevi08}%
  \BibitemOpen
  \bibfield  {author} {\bibinfo {author} {\bibfnamefont {P.~E.}\ \bibnamefont
  {Trevisanutto}}, \bibinfo {author} {\bibfnamefont {C.}~\bibnamefont
  {Giorgetti}}, \bibinfo {author} {\bibfnamefont {L.}~\bibnamefont {Reining}},
  \bibinfo {author} {\bibfnamefont {M.}~\bibnamefont {Ladisa}}, \ and\ \bibinfo
  {author} {\bibfnamefont {V.}~\bibnamefont {Olevano}},\ }\bibfield
  {booktitle} {\emph {\bibinfo {booktitle} {Ab Initio $GW$ Many-Body Effects in
  Graphene}},\ }\href {\doibase 10.1103/PhysRevLett.101.226405} {\bibfield
  {journal} {\bibinfo  {journal} {Phys. Rev. Lett.}\ }\textbf {\bibinfo
  {volume} {101}},\ \bibinfo {pages} {226405} (\bibinfo {year}
  {2008})}\BibitemShut {NoStop}%
\bibitem [{\citenamefont {Shao}\ \emph {et~al.}()\citenamefont {Shao},
  \citenamefont {Rudenko}, \citenamefont {Hu}, \citenamefont {Sun},
  \citenamefont {Zhu}, \citenamefont {Moon}, \citenamefont {Millis},
  \citenamefont {Yuan}, \citenamefont {Lichtenstein}, \citenamefont {Smirnov},
  \citenamefont {Mao}, \citenamefont {Katsnelson},\ and\ \citenamefont
  {Basov}}]{Shao19}%
  \BibitemOpen
  \bibfield  {author} {\bibinfo {author} {\bibfnamefont {Y.}~\bibnamefont
  {Shao}}, \bibinfo {author} {\bibfnamefont {A.~N.}\ \bibnamefont {Rudenko}},
  \bibinfo {author} {\bibfnamefont {J.}~\bibnamefont {Hu}}, \bibinfo {author}
  {\bibfnamefont {Z.}~\bibnamefont {Sun}}, \bibinfo {author} {\bibfnamefont
  {Y.}~\bibnamefont {Zhu}}, \bibinfo {author} {\bibfnamefont {S.}~\bibnamefont
  {Moon}}, \bibinfo {author} {\bibfnamefont {A.~J.}\ \bibnamefont {Millis}},
  \bibinfo {author} {\bibfnamefont {S.}~\bibnamefont {Yuan}}, \bibinfo {author}
  {\bibfnamefont {A.~I.}\ \bibnamefont {Lichtenstein}}, \bibinfo {author}
  {\bibfnamefont {D.}~\bibnamefont {Smirnov}}, \bibinfo {author} {\bibfnamefont
  {Z.~Q.}\ \bibnamefont {Mao}}, \bibinfo {author} {\bibfnamefont {M.~I.}\
  \bibnamefont {Katsnelson}}, \ and\ \bibinfo {author} {\bibfnamefont {D.~N.}\
  \bibnamefont {Basov}},\ }\bibfield  {booktitle} {\emph {\bibinfo {booktitle}
  {Electronic Correlations in Nodal-line Semimetals}},\ }\href@noop {}
  {\bibinfo  {journal} {(unpublished)}\ }\BibitemShut {NoStop}%
\bibitem [{\citenamefont {Rooney}\ \emph {et~al.}(2017)\citenamefont {Rooney},
  \citenamefont {Kozikov}, \citenamefont {Rudenko}, \citenamefont {Prestat},
  \citenamefont {Hamer}, \citenamefont {Withers}, \citenamefont {Cao},
  \citenamefont {Novoselov}, \citenamefont {Katsnelson}, \citenamefont
  {Gorbachev},\ and\ \citenamefont {Haigh}}]{Rooney}%
  \BibitemOpen
\bibfield  {journal} {  }\bibfield  {author} {\bibinfo {author} {\bibfnamefont
  {A.~P.}\ \bibnamefont {Rooney}}, \bibinfo {author} {\bibfnamefont
  {A.}~\bibnamefont {Kozikov}}, \bibinfo {author} {\bibfnamefont {A.~N.}\
  \bibnamefont {Rudenko}}, \bibinfo {author} {\bibfnamefont {E.}~\bibnamefont
  {Prestat}}, \bibinfo {author} {\bibfnamefont {M.~J.}\ \bibnamefont {Hamer}},
  \bibinfo {author} {\bibfnamefont {F.}~\bibnamefont {Withers}}, \bibinfo
  {author} {\bibfnamefont {Y.}~\bibnamefont {Cao}}, \bibinfo {author}
  {\bibfnamefont {K.~S.}\ \bibnamefont {Novoselov}}, \bibinfo {author}
  {\bibfnamefont {M.~I.}\ \bibnamefont {Katsnelson}}, \bibinfo {author}
  {\bibfnamefont {R.}~\bibnamefont {Gorbachev}}, \ and\ \bibinfo {author}
  {\bibfnamefont {S.~J.}\ \bibnamefont {Haigh}},\ }\bibfield  {booktitle}
  {\emph {\bibinfo {booktitle} {Observing Imperfection in Atomic Interfaces for
  van der Waals Heterostructures}},\ }\href {\doibase
  10.1021/acs.nanolett.7b01248} {\bibfield  {journal} {\bibinfo  {journal}
  {Nano Lett.}\ }\textbf {\bibinfo {volume} {17}},\ \bibinfo {pages} {5222}
  (\bibinfo {year} {2017})}\BibitemShut {NoStop}%
\end{thebibliography}%

\clearpage

\onecolumngrid
\begin{center}
\textbf{\large Supplemental Material: Electron-phonon interaction and zero-field charge carrier transport in the nodal-line semimetal ZrSiS}
\end{center}
\onecolumngrid

\setcounter{equation}{0}
\setcounter{figure}{0}
\setcounter{table}{0}
\makeatletter
\renewcommand{\theequation}{S\arabic{equation}}
\renewcommand{\thefigure}{S\arabic{figure}}

In this Supplemental Material, we present information regarding the role of spin-orbit coupling in electronic and vibrational properties of ZrSiS, and also provide convergence
behavior of Brillouin-zone integrals with respect to ${\bf k}$- and ${\bf q}$-space sampling, justifying numerical parameters used in the main text.

\subsection{Role of spin-orbit coupling in electronic and vibrational properties of ZrSiS}

Table \ref{table1} shows Raman active phonon frequencies in ZrSiS calculated with and without spin-orbit coupling (SOC), demonstrating the effect of spin-orbit coupling for zone center 
phonons.
The symmetry of the $\Gamma$ point coincides with the full point group symmetry of the lattice in real space, which is $D_{4h}$ in case of ZrSiS. Therefore, zone center optical 
phonons can be classified according to irreducible representations of the $D_{4h}$ point group as follows: $\Gamma_{D_{4h}} = 3E_g + 2E_u + 2A_{1g} + 2A_{2u} + B_{1g}$. Accordingly,
phonon modes containing $\emph{gerade}$ (\emph{ungerade}) symmetries can be characterized as Raman (infrared) active. For brevity, here we show Raman active frequencies only. 
From Table \ref{table1}, one can see that the relative frequency difference 
$\Delta \omega=(\omega_{\Gamma}-\omega^{\mathrm{SO}}_{\Gamma})/\omega_{\Gamma}$ is less than 0.1 \% in all cases. We conclude, therefore, that spin-orbit coupling has a negligible 
effect in the vibrational properties of ZrSiS.

\begin{figure}[b]
	\centering
    \includegraphics[width=0.75\textwidth]{{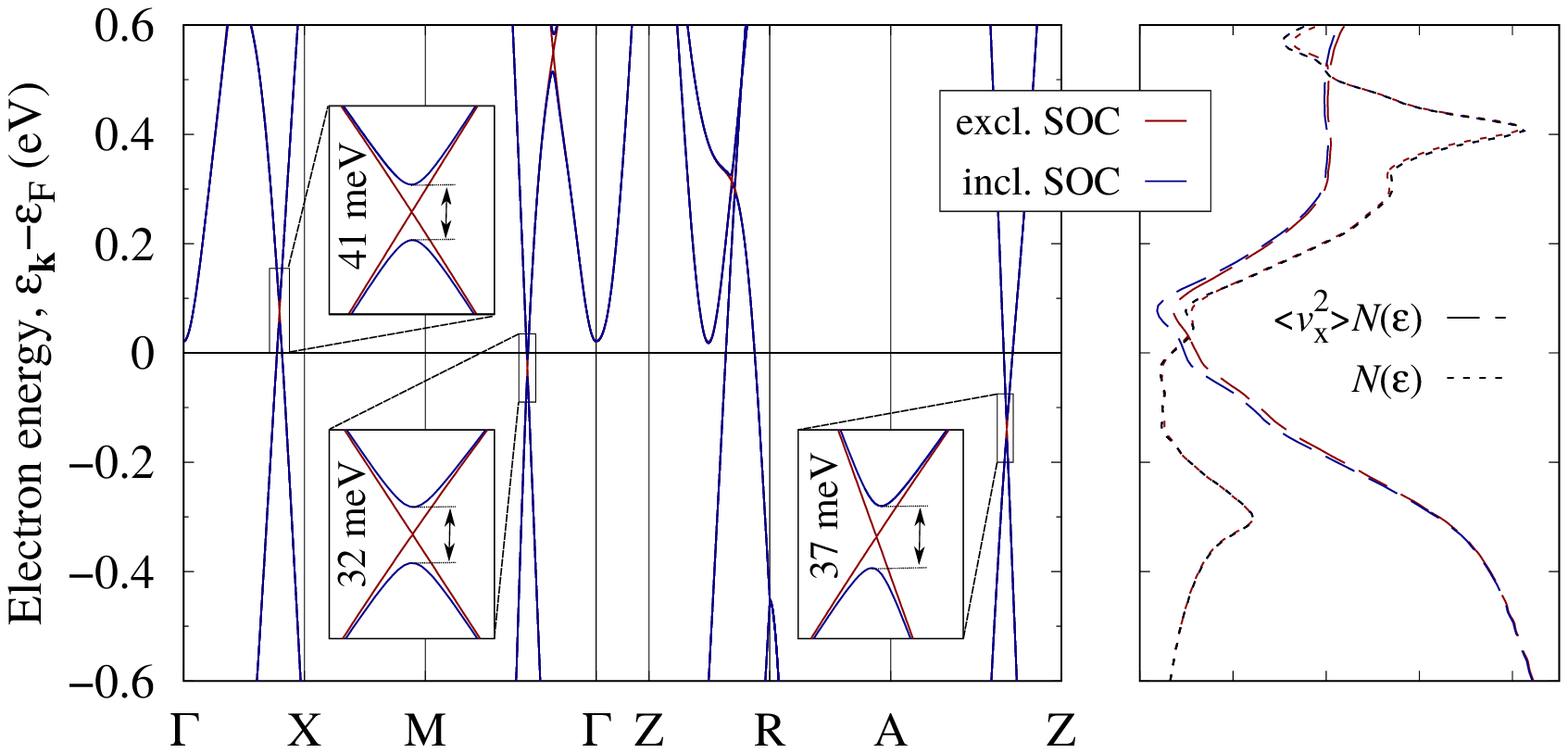}}
	\caption{Left panel: Band structure of ZrSiS calculated with and without SOC. The insets show a zoom-in on the band crossing region, where SOC-induced
gap opens up. Right panel: DOS $N(\varepsilon)$ and DOS-weighted square of the in-plane carrier velocity $\langle v_x^2 \rangle N(\varepsilon)$
in ZrSiS calculated as a function of energy with and without SOC (arbitrary units are used).}
	\label{bands_soc}
\end{figure}

Fig.~\ref{bands_soc} shows band structure, DOS, as well as Fermi-surface averaged $x$-component of the squared carrier velocity in ZrSiS calculated with and without SOC. From the band structure one can see that SOC induces a small gap at the band crossing points in the range 30--40 meV. This gap has practically no influence on DOS 
in the vicinity of the Fermi energy, but leads to a small reduction of the carrier velocities. By absolute values, the velocity reduction is below 3\% at the charge neutrality 
point ($\varepsilon=\varepsilon_F$). In the $z$-direction, the effect is even smaller (not shown here). Therefore, in the context of dc-transport, we observe
no significant influence of SOC on the electronic properties of ZrSiS.

\begin{table}[b]
        \caption{Raman active phonon frequencies (in cm$^{-1}$) in ZrSiS calculated with ($\omega_{\Gamma}$) and without ($\omega^{\mathrm{SO}}_{\Gamma}$) spin-orbit coupling,
shown in accordance with irreducible representations (irrep) of the $D_{4h}$ point group.} 
        \begin{tabular*}{1.0\linewidth}{@{\extracolsep{\fill}}cccccccccccc}
                \hline \hline
       Irrep                                       &  $E_g$ &    $A_{1g}$ &  $A_{1g}$ &  $B_{1g}$ & $E_{g}$ & $E_{g}$ \\ \hline
      $\omega_{\Gamma}$                            &  139.27 &   224.12   &  311.95   &  312.40 &   345.09  &   385.40     \\
      $\omega^{\mathrm{SO}}_{\Gamma}$              &  139.24 &   224.21   &  311.97   &  312.49 &   345.14  &   385.52     \\
                \hline\hline
        \end{tabular*}
        \label{table1}
\end{table}

\subsection{Convergence of Brillouin-zone integrals}

\begin{figure}[b]
	\centering
    \includegraphics[width=0.90\textwidth]{{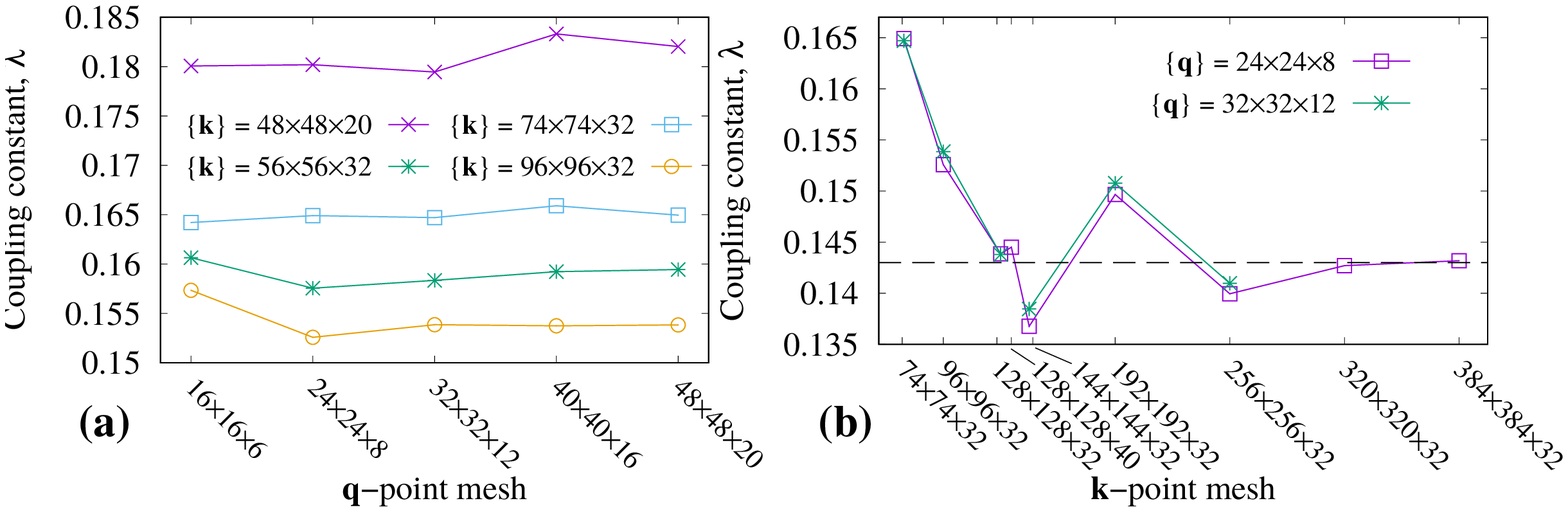}}
	\caption{Convergence behavior of the electron-phonon coupling constant $\lambda$ shown with respect to various ${\bf k}$- and ${\bf q}$-meshes used for numerical integration.
 Dashed line shows the value presented in the main text, which is considered as converged.}
	\label{lambda_conv}
\vspace{1cm}
	\centering
    \includegraphics[width=0.90\textwidth]{{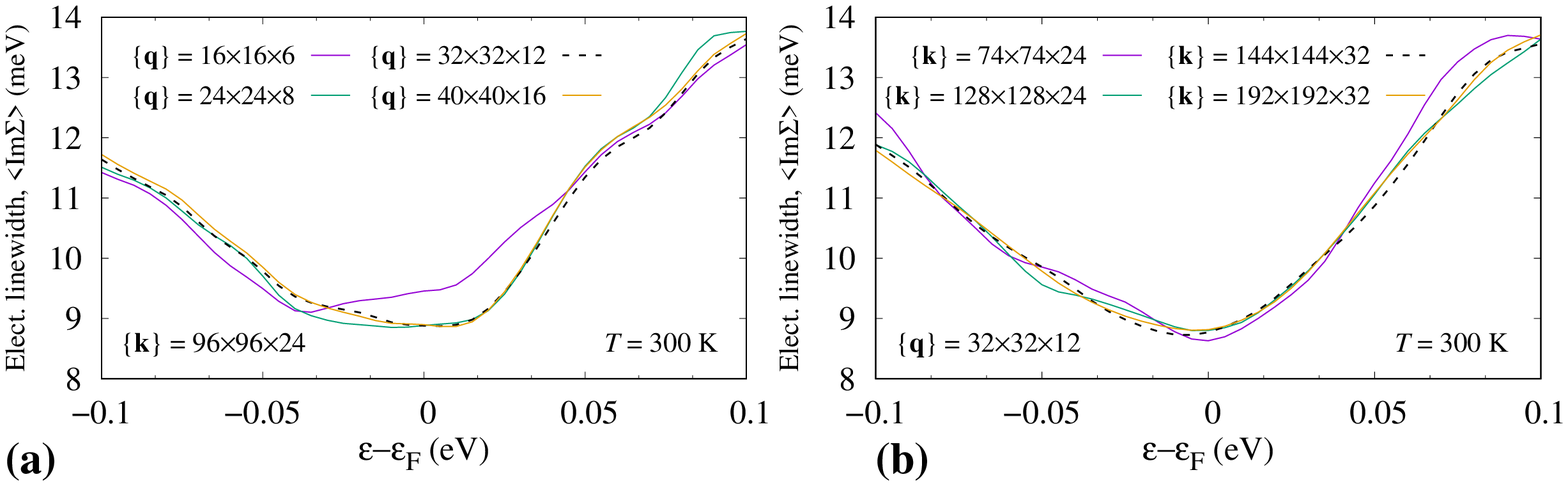}}
	\caption{Convergence behavior of the averaged electron linewidth $\langle \mathrm{Im}\Sigma \rangle$ shown as a function of the Fermi energy for various ${\bf k}$- and 
${\bf q}$-meshes. Dashed line in (a) and (b) corresponds to the result, considered as converged with respect to ${\bf q}$- and ${\bf k}$-meshes, respectively. $T=300$ K is used
in all cases presented.}
	\label{sigma_conv}
\end{figure}

Here, we demonstrate convergence behavior of relevant numerical integrals with respect to the density of {\bf q}- and {\bf k}-point meshes. Respectively, we consider 
(i) integrated electron-phonon coupling constant $\lambda$, which is a quantity related to the phonon linewidth, $\lambda \sim \sum_{{\bf q}\nu} \mathrm{Im}[\Pi_{{\bf q}\nu}]$; 
and (ii) Fermi-surface averaged electron linewidth, $\langle \mathrm{Im} \Sigma \rangle_{\varepsilon} \sim \sum_{n{\bf k}}\mathrm{Im}[\Sigma_{n{\bf k}}] \delta(\varepsilon-\varepsilon_{n{\bf k}})$.
Fig.~\ref{lambda_conv} shows $\lambda$ calculated using a number of different ${\bf q}$- and ${\bf k}$-point meshes. One can see that an accurate sampling of the ${\bf k}$-space is considerably
more important compared to the ${\bf q}$-space sampling. From Fig.~\ref{lambda_conv} it can be concluded that the density of ${\bf k}$- and ${\bf q}$- meshes used in the main text to 
calculate quantities related to the phonon linewidth [($384\times384\times32$) and ($24\times24\times8$), respectively] are sufficient to reach the numerical accuracy.

Fig.~\ref{sigma_conv} demonstrates convergence behavior of the averaged electron linewidth $\langle \mathrm{Im}\Sigma \rangle$. Similar to the electron-phonon coupling constant,
$\langle \mathrm{Im}\Sigma \rangle$ is less sensitive to the ${\bf q}$-mesh, and can be considered as converged if a ($32\times32\times12$) mesh is used. 
Again, ${\bf k}$-point sampling is more decisive to accurately describe the properties related to the electron self-energy. An essentially converged values of 
$\langle \mathrm{Im}\Sigma \rangle$ are obtained using a ($144\times144\times32$) ${\bf k}$-point mesh, which is adopted to obtain the results presented in the main text.

\end{document}